\documentclass[prd,twocolumn,showpacs,showkeys]{revtex4}
\usepackage{amsmath} \usepackage{bm} \usepackage{graphicx}
\newcommand{\bra}[1]{\langle#1|} \newcommand{\ket}[1]{|#1\rangle}
 \newcommand{\De}{\Delta}
\newcommand{\la}{\lambda} \newcommand{\be}{\begin{equation}}
\newcommand{\ee}{\end{equation}} \newcommand{\hc}{^\dagger}
\newcommand{\Hkan}{H^\Delta}
\newcommand{\Heff}{H_\lambda} \newcommand{\bmnie}{}

\begin{document}
\title{Large-momentum convergence of Hamiltonian bound-state
dynamics of effective fermions in quantum field theory} 
\date{April 19, 2002} 
\author{Stanis{\l}aw D. G{\l}azek} \email{stglazek@fuw.edu.pl}
\author{Marek Wi{\c e}ckowski} \email{wiecko@fuw.edu.pl}
\affiliation{Institute of Theoretical Physics, Warsaw University,
ul. Ho\.za 69, 00-681 Warsaw, Poland}
\preprint{IFT/13/02}

\begin{abstract}
Contributions to the bound-state dynamics of fermions in local quantum 
field theory from the region of large relative momenta of the constituent 
particles, are studied and compared in two different approaches. The 
first approach is conventionally developed in terms of bare fermions, a 
Tamm-Dancoff truncation on the particle number, and a momentum-space 
cutoff that requires counterterms in the Fock-space Hamiltonian. The 
second approach to the same theory deals with bound states of effective 
fermions, the latter being derived from a suitable renormalization group 
procedure. An example of two-fermion bound states in Yukawa theory, 
quantized in the light-front form of dynamics, is discussed in detail. 
The large-momentum  
region leads to a buildup of overlapping divergences in the bare Tamm-Dancoff 
approach, while the effective two-fermion dynamics is little influenced by the
large-momentum region. This is illustrated by numerical estimates of the 
large-momentum contributions for coupling constants on the order of between 
0.01 and 1, which is relevant for quarks.
\end{abstract}
\maketitle
\section{Introduction}
The notion of a bound state of fermions is based mainly on the
examples of  atoms and nuclei. The common feature of these systems is
that they are  non-relativistic. This means three things. Kinetic
energies of the fermions  are small in comparison to their rest mass
energy. Dominant interactions are  not able to produce fast moving
fermions from the slow ones and hence no  significant large-momentum
spin effects are generated. Creation of additional  particles can be
neglected and one can describe the bound states as built  from a fixed
number of fermion constituents. These features are all related to the 
fact that the domain of large relative momenta is not important, which
is easiest to describe in mathematical terms in the case of bound states 
of two fermions, such as positronium or deuteron. Their wave functions 
are self-consistent  solutions to the non-relativistic Schr\"odinger
equation $H|\Psi\rangle = E  |\Psi\rangle$ with Hamiltonian $H = H_0 +
H_I$, where $H_0$ denotes the  kinetic energy operator and $H_I$
stands for the interaction operator. The  matrix element
$\langle12|H_I|1'2'\rangle$ is the quantum Coulomb, or  Yukawa
potential with a repulsive core, respectively. The ket $|12\rangle$
denotes a state of two fermions labeled 1 and 2, with all their
quantum numbers collectively denoted by these labels. The
self-consistency of this well known  picture means that the wave
function $\psi(1,2) = \langle 12|\Psi \rangle$  quickly vanishes when
the relative momentum of fermions in the bound-state rest frame of
reference, $\vec p = \vec p_1 - \vec p_2$, approaches values
comparable with the fermions' reduced mass $\mu$ (the speed of light,
$c=1$).  

The success of the Schr\"odinger picture for bound states of
fermions extends  also to quarks. This is reflected in the constituent
quark model (CQM) being  the primary means for classification of
hadrons in the particle data tables,  and providing a benchmark for
more advanced approaches. However, the self-consistency  of the
non-relativistic picture is much harder to maintain for bound states 
of quarks up, down,  and strange, than for systems such as positronium 
and deuteron. This is because
the hadronic wave functions tend to have considerable components with
$|\vec p\,|$ comparable to or  exceeding $\mu$ for the quarks, and the 
domain of large relative momenta begins to play a significant role
in the binding mechanism. One is also interested in the description of  
hadrons moving with speeds very close to the speed of light. Since the 
fast  moving hadrons and their interactions cannot be consistently 
described within  a purely non-relativistic Schr\"odinger framework, 
theorists use the Feynman parton model in that case. Unfortunately, 
the binding mechanism of partons is  far beyond the scope of the 
parton model. As alternative to these models, one  can approach the 
issue of bound states of fermions using quantum field theory  (QFT), 
where the corresponding operator $H$ appears to contain all the 
relevant information about relativistic effects in the domain
of large relative momenta of the constituents.

However, the relativistic description of bound states of fermions in
QFT,  especially in quantum chromodynamics (QCD) in the case of light
quarks,  makes the conceptual difficulties with the constituent
picture even greater  than in the simple models. In the equation
$H_{QCD} |\Psi\rangle = E|\Psi\rangle$, all factors remain unknown.
Major reasons for this status of the theory originate in the 
large relative momentum domain in the motion of virtual particles.
This is illustrated by the following examples. One  is that local
QFTs lead to canonical interaction Hamiltonians $H_I$ that change
individual bare particle energies by unlimited amounts (spin-dependent
factors grow with momentum transfers). The large momentum range is
enhanced and leads to divergences, invalidating the concept of a
non-relativistic picture  entirely unless special conditions, such as
an extraordinarily small coupling, are met. Another example is that the
interactions create new bare particles and this  effect contributes to
the boosting of bound states, which implies that the  motion of bound
states is associated with multi-particle components and the {\em ad hoc}
limitation to a fixed number of bare constituents goes out the window
in relativistic QFTs. A third example is that even the state with 
no constituent particles, i.e. the ground state of a theory, or 
vacuum, proves to be so complicated that no approximate solution 
of verifiable accuracy has been conceived yet,
although many ansatzes have  to be and are employed in practical
attempts. In these circumstances, the main theoretical approach to 
bound states of quarks (mainly heavy ones that move slowly) is based 
on the lattice  version of QCD, and great progress has been achieved 
in numerical studies of  the theory that way. Nevertheless, it appears 
that a quantitative explanation of how the constituent picture with 
a simple Hamiltonian could be an approximate solution, remains a conceptual 
and quantitative mystery. No detailed constituent  wave function picture
for relativistic field quanta in Minkovsky space has been theoretically 
identified or is expected to readily follow from the lattice approach 
alone. The question of convergence of the binding mechanisms in the 
domain of large relative momenta of constituent particles, remains open. 

This article provides some numerical arguments that the required
constituent picture with well-controlled large relative momentum
domain  may become in principle identifiable if one provides a
precise definition  of the constituents as effective particles, in
distinction  from the  bare quanta of the local theory. 
Thus, the process of solving a theory is arranged in two steps, which
is typical in lattice approach \cite{lattice}, or sum rules \cite{SVZ}.
In the first step, one derives an effective dynamics, and, in the second
step, one attempts to solve the effective theory instead of dealing 
directly with the original degrees of freedom. Here, one derives
the effective fermions  of size $\lambda^{-1}$ using a suitable
boost-invariant perturbative renormalization  group procedure for
their Hamiltonians. The procedure is carried out up to
second-order perturbation theory, and the resulting dynamics  is
compared with the standard picture, where the finite scale $\lambda$ is
absent. In distinction from the diverging bare dynamics, the
effective  one comes out limited to the momentum range given by $\lambda$, and
this scale is  reduced using differential equations to the most 
suitable one for description of bound state properties  in terms 
of a fixed number of
the corresponding constituents. In the renormalized  Hamiltonian
picture, the point-like bare particles of the local theory correspond
to $\lambda = \infty$ and their dynamics heavily involves large
momenta, and  multi-particle states, for any finite value of the
coupling constant. However,  the situation is completely changed when
$\lambda$ is lowered to values comparable to the bound state
masses. The binding is described by a new  Schr\"odinger equation,
$H_\lambda  |\Psi\rangle = E |\Psi\rangle$, where the  Hamiltonian,
$H_\lambda = H_{0\lambda} + H_{I\lambda}$, is written in terms of
creation and annihilation operators for the effective particles,
$b^\dagger_ \lambda$ and $b_\lambda$ for fermions, $d^\dagger_\lambda$
and $d_\lambda$  for anti-fermions (and $a^\dagger_\lambda$ and
$a_\lambda$ for bosons). This effective particle picture is discussed 
in the present paper.

The key physical reasons for the hope that the effective constituent
picture  does emerge from QFT can be understood by recalling again what
happens in the well-known cases of atoms (or positronium) and nuclei 
(or deuteron) discussed
above. These systems can be  understood in terms of constituents for
quite different reasons. The  explanation of the difference is limited
below to the positronium and deuteron,  but the two examples are
sufficient to make the point that concerns  all bound systems
of fermions in QFT. In the Schr\"odinger quantum-mechanical  picture
of positronium, the coupling constant that appears in the Coulomb
potential is very small in comparison to 1, i.e.  $e^2/4\pi=\alpha \sim
1/137$. Therefore, the interaction produces quite small binding
energy, of order $\alpha^2 \mu/2$, and the $e^+e^-$ bound-state mass
is dominated by $2 m_e$. The relative-motion wave function is
proportional to $(\alpha^2 \mu^2 + |\vec p\,|^2) ^{-2}$, independently 
of the fermion spins. When one extends this picture by embedding it
in quantum electrodynamics (QED), one finds out that the initial wave
function is so small for large momenta $\vec p$, that no significant
correction is able to emerge from that region and alter the original
picture  with the Coulomb potential. This is found by expanding the
theory term by term  in powers of $\alpha$ around the initial
Schr\"odinger picture. The interaction  linear in $\alpha$ (Coulomb
force) is sufficient to describe the main features  of fermionic bound
states in QED, and higher powers of $\alpha$ are not important  for
theoretical understanding of the bulk of the bound state
structure. Although  the integrals in the corrections run over the
momentum range that formally  extends far beyond $\mu$, the coupling
constant is too small for the relativistic  fermion spin factors and
particle creation processes to produce any major  modifications of the
leading approximation. This feature will be farther  discussed in the
next Sections.

In the meson-exchange models of the deuteron binding mechanism, the
analogous  coupling constant is three orders of magnitude larger than
in QED. If one  attempted to use QFT to derive the Yukawa potential
using the same strategy  as in QED, and to calculate corrections, the
perturbative procedure would fail.  The interactions would accelerate
nucleons to the speed of light almost  immediately on the bound-state
formation time-scale,  many new particles would be created, and
the large momentum relativistic ``corrections'' would dominate the 
``leading'' non-relativistic terms. One could then ask why the famous
one-boson-exchange (OBE) potentials, such as the Yukawa potential
with a core,  could still be used in phenomenology of relativistic
nuclear physics and work  self-consistently in the nuclear bound-state
equations anyway. What saves the  picture of a fixed number of
relatively slow nucleons interacting through  exchange of mesons from
serious inconsistency when one includes the elements of QFT, is that 
the interactions  responsible for
emission and absorption of mesons by nucleons contain form  factors
that limit momentum transfers to values so small that the interactions
are effectively weak, much weaker than a change of $\alpha$ in QED by
the  factor 1000 would imply. Consequently, the binding energy 
is  much smaller than the sum of two nucleon masses, e.g. about
2.2 MeV for deuteron. 
The wave functions  of such generalized relativistic nuclear
physics picture are not overwhelmingly extending into  the 
large-relative-momentum
domain because the form factors eliminate coupling  to that region and
the non-relativistic Yukawa potential with a repulsive core  is not
invalidated by huge corrections. It could not be so with bare
point-like  fermions in local QFT, but it does work in the
phenomenological picture of  effective particles with the form
factors. By the way, this example is not  intended to suggest that
nucleon dynamics should be completely derivable  directly from a local
QFT that ignores the existence of quarks. In fact, a  scenario for how
to derive the effective nuclear physics picture from QCD will  be
discussed in the last Section of this work. Nevertheless, the nuclear
physics  picture does indicate that an effective particle dynamics may
involve large  coupling constants in potentials that resemble
perturbative second-order  interactions with form factors.

In QCD, none of the two schemes can apply separately. On the one hand,
the  effective coupling constant in the constituent QCD picture cannot
be as  small as in QED, because QCD is characterized by asymptotic
freedom, or  infrared slavery. This means that the effective coupling
strength is expected  to grow when the scale of relevant momentum
transfers decreases. The coupling  constant is already on the order of
0.1 at transfers on the order of 100 GeV  and it may be much larger
for transfers on the order of nucleon masses.  Therefore, the effects
that have marginal size in the eigenvalue problem for  approximate
$H_{QED}$, such as spin splittings, are expected to be much larger
and more important  for
understanding eigenstates of $H_{QCD}$, and the initial approximation
is  not as simple as in QED. On the other hand, one cannot freely
insert form  factors into the local Lagrangian for quark and gluon
fields because it would  spoil the local gauge symmetry structure. The
contact with QCD would be  irreversibly lost.

The situation is changed when QFT is approached using the idea of
similarity  renormalization group procedure for Hamiltonians
\cite{sim1}, especially in  the case of QCD \cite{KWetal}, and when
one combines the similarity idea with  the concept of form factors in
the Hamiltonian interaction vertices for  effective particles
\cite{Gacta}. Initially, the coupling constant is small due to
asymptotic freedom and one can think of using the small coupling
constant  in canonical QFT to solve the renormalization group
equations for $H_\lambda$  using a perturbative expansion. The method
avoids small energy denominators in  the perturbative calculation
entirely and the non-perturbative part of the  dynamics remains
untouched in the perturbative calculation of $H_\lambda$. That way 
one derives  effective-particle Hamiltonians that involve vertex 
form factors
of small  width $\lambda$ in the interaction terms. One can have
sizable couplings  in $H_\lambda$ with small $\lambda$, as required
by infrared slavery, without  losing control over the size of
corrections to the leading constituent picture  in diagonalizing
$H_\lambda$. The spectrum of such Hamiltonians can be sought
numerically because the form factors limit the range of momenta
strongly enough  for a discrete computer code to cover the pertinent
region, as in the nuclear physics case discussed above. This idea has
already been studied qualitatively in a simple numerical matrix model
\cite{GWafbs} using Wegner's flow equation  \cite{Wegner} (and its
generalizations). The more detailed effective particle  calculus used
in the present work in the case of Yukawa theory, is already  known to
produce asymptotic freedom in $H_\lambda$ for QCD \cite{Ggluon}.  The
new approach has been also extended to the whole Poincar\'e algebra
\cite{poincare} in QFT.

The present paper provides elementary numerical estimates of the
orders of  magnitude of the interactions that appear in QFT in the
bound state dynamics  of two effective fermions of size
$\lambda^{-1}$. Yukawa theory is used to avoid  complications related to
gauge symmetry while one still preserves some of the  singular 
large-momentum components in the spinor factors that characterize
fermions. The well-known issue of triviality in Yukawa theory is 
irrelevant here since our goal is to estimate the size of
corrections in the bound-state dynamics in an effective theory,
rather than in the ultra-violet (UV) dynamics of the initial QFT. 
The Yukawa example serves only as a source of typical UV factors
that QFTs provide anyway, no matter if the theory is trivial,
asymptotically free, or otherwise.

The key qualitative question is by how much the Hamiltonian
$H_\lambda$  derived in QFT might differ from familiar models,
especially from the  non-relativistic Schr\"odinger picture with a
Yukawa potential (or a Coulomb  potential in the case of exchange of
massless mesons), for given values of  $\alpha$ and $\lambda$. Another
question is related to the fact that the exact  solution of
renormalization group equations for $H_\lambda$ and subsequent  exact
diagonalization of $H_\lambda$ should lead to spectra that are
independent  of $\lambda$. However, when one calculates $H_\lambda$ in
perturbation theory  of low order, such as the second order that
characterizes the Coulomb and  Yukawa potentials, the
$\lambda$-dependence must appear. Bound-state energies  may depend on
$\lambda$ when $\lambda$ is made too small or $\alpha$ is made  too
large. The question is how large is the residual $\lambda$ dependence
of  second-order corrections to the Coulomb-like picture in how large
a range of  values of $\alpha$ and $\lambda$ that can be
self-consistently (i.e., without  significant $\lambda$-dependence)
reached in lowest orders of perturbation  theory, for $H_\lambda$
itself in the renormalization group part of the calculation, and 
for the eigenvalues and wave functions in
the bound state perturbation theory around the Coulomb-like
ansatz. Both questions are answered in the next Sections by
estimating the  size of those corrections that are most important for
large momenta, and which would  lead to divergences in the absence of
$\lambda$. The results imply that the  most dangerous corrections that
might diverge in the absence of $\lambda$, turn  out to be quite small
even for sizable coupling constants. This provides some initial
ground for the hope that the effective particle dynamics can be
attempted numerically in QCD in the same way. An outline of how the
same approach, if successful, could then in principle also lead from
effective $H_{\lambda QCD}$  to the effective meson-exchange
interactions between nucleons in nuclear physics,  is also provided.

Many farther dynamical issues of great interest, concerning effective
fermions  and their binding in QFT, can be posed and investigated using
the effective  particle calculus, but none are studied here in
detail. For example, the case  of phonon exchange between electrons as
approximated by a QFT is not discussed  here \cite{WPR}, and symmetry
restoration issues in not-explicitly symmetric  Hamiltonian approach
through coupling coherence \cite{PPR} is also not considered.
Instead, this work is focused on the readily accessible quantitative
estimates  that show the magnitude of the difference between the
convergent two-effective  fermions bound-state dynamics and the
diverging dynamics of two bare  fermions in the approaches
based on the Tamm-Dancoff truncation in local QFT  \cite{tamm,
dancoff}. In fact, this paper starts with the Tamm-Dancoff-like
approach to local theory, because it is more familiar. Then, the
effective  particle approach is introduced, with its new features
exposed through the  contrast with the Tamm-Dancoff one. Thus, Section
\ref{sec:def} provides  definitions of the renormalized Tamm-Dancoff
scheme in Subsection \ref{sub:RTDa}, and the effective particle scheme
in Subsection \ref{sub:Repa}. Both approaches  involve a universal
procedure for obtaining a two-fermion eigenvalue equation that can be 
compared with the non-relativistic
Schr\"odinger equation. This universal procedure is for brevity called
{\it  reduction}. It is denoted by the symbol $R$ and described in
subsection \ref{sub:R}.  Section \ref{sec:can} introduces a bare
light-front Hamiltonian in Yukawa quantum  field theory that serves as
a starting point for subsequent Sections. The canonical  Hamiltonian
is supplied with some regularization factors $r_\De$ and counterterms.
Section \ref{sec:A1} describes details of the approach that treats
bound states  of fermions as if they could be viewed as made
of two bare fermions.  This approach is shortly called {\it approach
1} and encounters large-momentum conceptual and  calculational 
difficulties that are removed by switching over to the approach  
discussed in detail in Section \ref{sec:A2}. Namely, Section \ref{sec:A2} 
presents  the effective fermion approach, called {\it approach 2}, 
using comparison and contrast with the  approach 1. 
In the approach 2, bound states of 
fermions are treated as built  from effective fermions of size
$\lambda^{-1}$. Conclusions are drawn in Section  \ref{sec:con}. Appendixes
contain information about the operation $R$, spinor factors, momentum 
variables, and a Coulomb bound state wave function used in the 
calculations.

\section{Definitions}
\label{sec:def}

This Section provides definitions of two light-front Hamiltonian
approaches to the bound state dynamics, the renormalized Tamm-Dancoff
approach (approach 1), and effective particle approach (approach 2).
They are compared through numerical estimates in next Sections in the
case of a bound state of two fermions. The procedure of reducing  a
theory to the two-fermion sector is mathematically the same in  both
cases, except for a different definition of what is meant by  the
two-fermion states. In the approach 1, the starting point will  be
identical with bare fermions in QFT, in the approach 2 it will  be the
effective ones. The reduction procedure, denoted by $R$, is
described at the end of this Section for both approaches
simultaneously.
 
\subsection{Renormalized Tamm-Dancoff approach}
\label{sub:RTDa}

This approach is represented by the following diagram,
\begin{equation}
{\cal L} \,\, \stackrel{0}{\rightarrow} \,\, :H_{can}: \,\,
\stackrel{i}{\rightarrow} \,\, :H_{can}^\Delta:+X^\Delta \,\,
\stackrel{ii}{\longleftrightarrow} \,\,  \rm{solve} \,\, .
\end{equation}

{\em 0)} The initial step on the left denotes a canonical  derivation
of a field theory Hamiltonian from its Lagrangian,  quantization, and
normal ordering with respect to the bare  vacuum state, dropping all
diverging terms on the basis of hindsight knowledge that the normal
ordered Hamiltonian will eventually contain counterterms of the same
structure, but well defined.

{\em i)} The next step is called regularization, which is  necessary
because the canonical Hamiltonian leads to ultraviolet  divergences due to
the infinite range of energy scales to sum  over in the intermediate
states when one calculates physical   observables (such as scattering
cross-section) using perturbative  formulae. One imposes an
ultraviolet cutoff (we denote it by $\De$)  on the range of momenta that
are included in the sums. Observables  calculated with such
regularized Hamiltonian explicitly depend on both the parameter
$\De$, and on the way the regularization is imposed. For example, the
regularization may preserve or violate  some symmetries - an oblate
cutoff function would not be  rotationally symmetric, or a
frame-dependent cutoff would not respect Lorentz symmetry, and finite 
corrections due to such regularizations would be required. To remove
the artificial dependence of  observables on regularization,  one 
has to add new terms to the Hamiltonian (called counterterms and 
denoted $X^\De$) that also depend on the regularization.  The construction 
of counterterms is based on the idea of recovering the contribution from 
dynamics above the cutoff, which is to lead to finite and cutoff 
independent results when  $\De \rightarrow \infty$. The counterterms
structure is thus  determined by the structure of divergences. The
latter can be computed and $X^\De$ is than adjusted to cancel the
cutoff dependence. It is  less straightforward to remove finite
effects of regularization, but this issue will not be important in the
next Sections. The regularized  Hamiltonian $:H_{can}^\Delta:+X^\Delta$ 
is denoted by $\Hkan$.
 
{\em ii)} The last arrow indicates solving of the eigenvalue equation
for $\Hkan$. A two-steps procedure is used.

{\em Step a)} First one finds eigenstates of $H^\Delta$ whose dominant 
component for
vanishingly small coupling constants is equal to one bare fermion.
These states represent what one could call a physical fermion.  The
solution is found from the eigenvalue equation for the whole $\Hkan$ by
reducing that equation  with the help of operation $R$ to an equivalent 
equation for the Fock component with one bare fermion.
If one requires the eigenvalue
to be finite, one has to include in $X^\De$ a mass counterterm of
a calculable form.

{\em Step b)} Then, one makes a reduction $R$ of the eigenvalue problem
for $H^\Delta$ to a two-bare-fermion
subspace, to find an eigenstate of the Hamiltonian $\Hkan$ that is
dominated by a pair of bare fermions for infinitesimally small
coupling constants. The parameters in the resulting two-fermion
eigenvalue problem are expressed in terms of the physical fermion mass
found in step {\em a)} above. It turns out that the 
calculated eigenvalues still depend on the cutoff (some diverge if
$\De\rightarrow\infty$), although the individual matrix elements of
the reduced two-body Hamiltonian do not depend on the cutoffs once one
includes mass counterterms calculated in {\em ii a}). Therefore, there
is a problem of how to construct counterterms that would remove $\De$
dependence from physical results \cite{pinsky}.

It is important to mention here an alternative approach, which is also
inspired by the Tamm-Dancoff truncation but includes renormalization
group idea in a more sensible way \cite{Perry:mz}. Namely, one can  
assume that some effective fermion representation of the eigenstates 
of $\Hkan$ does exist. One tries then to build a triangle of 
renormalization (the notion of the triangle in the case of similarity
renormalization group scheme for light-front Hamiltonians can be 
found in Ref.  \cite{KWetal}) for a sequence of calculations with growing 
cutoffs $\Delta$ and growing numbers of fermions and bosons, adjusting 
counterterms to the momentum cutoffs differently in different Fock sectors. 
The hope is to understand regularities in the obtained structures that 
produce stable outputs when the momentum cutoffs and the particle numbers 
are made so large in the triangle that one can follow the renormalization 
group flow down from the large values and recognize interactions that 
have a universal character. Moving then along the renormalized Hamiltonian 
trajectory in the triangle, one would be able to arrive at the desired 
effective particle 
picture. One of the key tools one would use  in studying the triangle 
would be the coupling coherence \cite{PPR} that  requires that the 
interactions retain their forms when one moves around  the triangle with 
large values of the cutoffs. This approach is thus relying  on regularities 
that need to be discovered in a vast  amount of intricate 
analytic and numerical data. The data can also be interpreted  with the help 
of perturbative techniques wherever possible. As a variation of this idea,
one may even attempt to include new degrees of freedom, such as Pauli-Villars 
pseudo-particles, and treat them  as a suitable way of writing down 
required counterterms \cite{zle}. However, the latter strategy faces a problem
of sending the regulator masses to infinity (it is not clear how to do it
using computers) or living with the additional degrees of freedom that 
are absent in the initial theory. 
One also uses discretization of the longitudinal light-front
momenta (usually denoted by $p^+$) \cite{casher}, which limits the number 
of constituents.
However, the transverse dynamics is 
not smoothed out that way and poses the same divergence problems. The 
initially helpful longitudinal cutoff has to be relaxed in order to recover 
boost invariance. At that stage, the same problems reappear fully again and 
remain unsolved. Thus, the already known options for general searches for 
effective fermion dynamics are hard to pursue. In contrast, in the present 
article, it is taken for granted that one can assume from the very beginning 
a large degree of rigidity in the derivartion of effective theory by writing
it in terms of creation and annihilation operators for effective
particles. The assumption allows one to narrow the search for
effective dynamics using a unitary
renormalization group procedure \cite{Gacta}, which is the basis for the 
approach 2 introduced below. 

\subsection{Renormalized effective particle approach}
\label{sub:Repa}

The procedure consists of three steps that are described later in
detail in Section \ref{sec:eff}, and can be represented by the
following diagram.
\begin{equation}
{\cal L} \,\,  \stackrel{0} \rightarrow \,\,  :H_{can}: \,\,
\stackrel{i} \rightarrow \,\,  :H_{can}^\Delta:+X^\Delta \,\,
\stackrel{U_\lambda}{\longleftrightarrow}\,\, H_\la
\stackrel{iii}\rightarrow \rm{solve}
\end{equation}

The steps {\em 0)} and {\em i)} are the same as before except that one
works with the bare creation and annihilation operators for efficient 
book-keeping for Hamiltonian terms at all times, instead of storing 
a huge number of selected matrix elements of $\Hkan$. Note that many other 
operators could have the same selected matrix elements. Also, matrix 
elements of operators change 
when the basis states change even if the operators remain unchanged, 
and the same operator may have different matrix elements when different 
states are used to represent constituent particles.

{\em ii)} This step, marked with $U_\lambda$ in the diagram, is made 
using the procedure of renormalization group for effective particles 
(RGEP) \cite{Gacta,Ggluon}. The procedure is defined as a unitary rotation 
of creation and annihilation operators by an operator $U_\la$. Hamiltonians 
$H_\la$ are expressed in terms of the effective-particle creation and 
annihilation operators that depend on the ``width'' parameter $\lambda$. 
$\la$~ranges from $\infty$ in $\Hkan$ to a finite value on the order of 
bound-state masses in the effective constituent dynamics. That such 
small scales can in principle be reached using perturbation theory 
\cite{GWafbs}, is possible because RGEP procedure is designed according 
to the rules of the similarity  renormalization group for 
Hamiltonians \cite{sim1}. It separates the perturbative from non-perturbative
aspects of the  theory (see the original articles). By the same
design, the Hamiltonian  $H_\la$ cannot change invariant masses of
effective-particle Fock states by more than about $\la$ in a single
interaction.
\begin{figure}[htbp]
\begin{center}
\includegraphics[width=6cm]{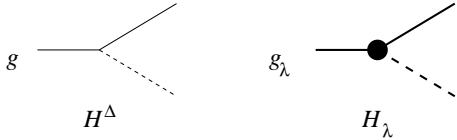}
\caption{The main difference between $H^\De$ and $H_\la$ is that the
latter is for effective particles of size $\la^{-1}$, instead of the
point-like bare ones that the former is for, and interaction terms 
in $H_\la$ are limited by form factors $f_\la$.  The form factors 
are marked in figures by a black blob.}
\label{fig:YvsfY}
\end{center}
\end{figure}
Thus, the emission of effective bosons by effective fermions is  possible
only if the associated kinetic energy change of relative motion of the 
particles does not exceed $\la$.
Consequently, when $\la$ is small, Fock sectors with different numbers
of effective particles are coupled weakly even for sizable coupling
constants, like in the nuclear physics example discussed in the
Introduction.

{\em iii)} This step is analogous to the step {\em ii)} in the
approach 1 and amounts to solving the eigenvalue problem for effective
Hamiltonian $H_\la$. The key difference, however, is that when one
works using the basis of effective particles in the Fock space, states with 
two effective fermions couple  only to states with similar relative momenta.
Therefore, the large relative momentum  domain remains suppressed, and it can
be described using perturbation theory without assuming that the coupling
constant is very small. Thus, when one solves the eigenvalue equation for
$H_\lambda$, one can introduce a  new perturbation theory for
the reduction operator $R$, expanding in powers of $H_{I\lambda}$.
This gives an equivalent Hamiltonian that
acts only in the dominant Fock space sectors. There are two steps to
do, as in the approach 1.

In the {\em step a)}, one first considers eigenstates dominated by one
effective fermion, which defines a physical mass of a physical fermion
in  the approach 2. Then, in the {\em step b)}, one finds an equation
describing bound states of two effective particles. The parameter $\la$
is a key to  the procedure. Its value determines whether derivation of
the effective  Hamiltonian $H_\la$ and its reduction by the operation
$R$ to a model  subspace Hamiltonian, denoted by $H_R$, is possible in
perturbation theory.  The smaller is $\la$ the simpler are the
approximate solutions for bound  states of effective fermions, in the
sense that they tend to reduce to the dominant effective Fock
sector. But if $\la$ is too small, the step  {\em ii)} of derivation of
$H_\la$ in perturbation theory loses accuracy (the perturbative integration
of renormalization group equations begins to significantly cut into 
the bound-state dynamics). Therefore, $\lambda$ cannot be lowered too far 
using perturbation theory for $H_\lambda$. The optimal choice of $\la$ is 
the one that combines the simplest perturbative expansion for $H_\la$ with
least complicated computer  diagonalization of $H_\la$. The main criterion 
for choosing the right  range for $\lambda$ is that the calculated observables 
are not  sensitive to variation of $\la$ over that range.

The final comment concerns Refs. \cite{BP,Jones}, where a different
concept  of a bound state calculus has been developed using coupling
coherence in second-order perturbation theory for Hamiltonian matrix
elements, also in the similarity scheme but without the constraint to 
a boost-invariant unitary rotation of creation and annihilation 
operators (from the bare to effective ones). In distinction 
from these works, the approach 2 is not based on the coupling coherence 
because no coherent structure is known {\em a priori} in the region of 
small $\la$,  far from the initial canonical structure. Instead, one 
uses a perturbative  expansion for the effective-particle renormalization 
group flow in terms of a suitably defined  coupling constant and tries 
to find out the relevant structures in a prescribed basis, in which the 
expansion in powers of the coupling constant may be extrapolable to 
its physical values. Here, the analysis of Yukawa theory is limited to 
numerical estimates of the size of only a few important terms of first 
and second order in a bound-state perturbation theory, which goes beyond
the renormalization group flow. In the flow itself, it is already
known that higher-order calculations are certainly possible even in much 
more complex theories (see e.g. \cite{Ggluon}). The effective bound-state 
dynamics is virtually unknown \cite{briprep}.

\subsection{Reduction procedure}
\label{sub:R}

The following scenario occurs several times in next Sections.
There is an eigenvalue equation for a Hamiltonian
$H = H_0 + H_I$,
\begin{equation}
H\ket{\psi}=E\ket{\psi}\;,
\end{equation}
which is too large to solve exactly on a computer in the sense that
the number of {\em a priori} important basis states is infinite. One looks 
then for an equivalent Hamiltonian that acts only in a limited 
subspace of states. One way of constructing the model subspace 
dynamics is to use the transformation $R$ \cite{Bloch,Wold}. Details of
the transformation are written in Appendix \ref{app:red}. The idea
is following. 

One denotes the projection operator on the chosen subspace of the whole 
Fock space $F$ by ${\hat P}$, and the projector on 
the complementary space, $1-{\hat P}$, by ${\hat Q}$. If the interaction 
Hamiltonian $H_I$ is small in the sense that it only weakly couples states 
from the subspace $\hat PF$ to states in the subspace $\hat QF$, then one can 
calculate an operator $R$ that produces certain eigenstates of $H$ from 
eigenstates of a new Hamiltonian $H_R$ that has eigenstates contained in 
the subspace $\hat PF$. The transformation $R$ leads to the following 
expression for the Hamiltonian $H_R$ acting in the subspace $\hat PF$, 
expanded in powers
of the interaction Hamiltonian $H_I$ (cf. Eq. \ref{app:hr}).
\begin{eqnarray}
\label{app:hrij}
\bra{i}H_R\ket{j}&=&\bra{i}\left( {\hat P}H{\hat P}+ \frac{1}{2} {\hat
P}H_I\frac{\hat Q}{E_j-H_0}H_I{\hat P}+\right.\nonumber\\
&&\left.\frac{1}{2}{\hat P}H_I\frac{\hat Q}{E_i-H_0}H_I{\hat P}+\dots
\right)\ket{j}\;.
\end{eqnarray}
Note that the Hamiltonian $H_R$ does not depend on the eigenvalues 
of $H$, but only on the eigenvalues of $H_0$. Namely, $H_0 \ket{i}
= E_i \ket{i}$. In particular, one can define $H_0$ in conjunction 
with the subspace $\hat PF$ so that $H_I = H - H_0$ is as weak as one can 
get simultaneously with preserving control over the spectrum of $H_0$. 
For example, in the case of two fermions, it is often useful to take 
$H_0$ equal to a sum of the kinetic energy and Coulomb potential 
operators when the fermions interact through emission and absorption 
of massless bosons with a small coupling strength. The strength of 
the coupling (size of $H_I$) may be small when the coupling constant 
itself is small, or when a form factor allows only small range of 
momentum transfers from the fermions to bosons and between the fermions.

\section{Canonical theory}
\label{sec:can}

We consider a theory of fermions of two kinds interacting through a 
Yukawa coupling with scalar bosons. The common starting point for 
all approaches to a two-fermion bound-state dynamics discussed in 
this paper is the Lagrangian:
\begin{equation}
\label{can:L}
{\cal L}=\bar \psi(i\partial\!\!\!\slash -m -g\phi)\psi+
\frac{1}{2}\partial_\mu\phi \partial^\mu\phi \;,
\end{equation}
where  $\bar \psi=\left(\bar \psi^{(1)}, \bar \psi^{(2)}\right)$. 
That is, $\cal L$ includes the kinetic terms for two kinds of 
fermions (both of mass $m$), the kinetic term for massless scalar 
bosons and a point-like (Yukawa) interaction of the fermions with 
bosons.

The canonical Hamiltonian is defined using light-front 
quantization \cite{Yan}, where the evolution of states in 
``time'' $x^+ = x^0 + x^3$ is generated by a three-dimensional 
integral of the energy-momentum-density-tensor component 
$T^{+-}$ over the light-front hyper-plane $x^+=0$. The 
Hamiltonian is given by
\begin{widetext}
\begin{equation}
\label{can:pminus}
P^-=\frac{1}{2}\int dx^-d^2x^\perp : \left[  -\phi
\partial^{\perp2}\phi +\sum_{i=1}^2\left( \bar
\psi_{m}^{(i)}\frac{-\partial^{\perp2}+m^2}{i\partial^+}\gamma^+
\psi_{m}^{(i)}+ 2g\phi \bar \psi_{m}^{(i)} \psi_{m}^{(i)} +g^2  \bar
\psi_{m}^{(i)}\phi \frac{\gamma^+}{i\partial^+}\phi \psi_{m}^{(i)}
\right)\right]:\;.
\end{equation}
\end{widetext}
$\psi_m$ are fermion fields that satisfy free Euler-Lagrange equations
with mass $m$. Quantum commutation relations for all fields are satisfied 
by replacing all fields at $x^+=0$, [$x^\perp = (x^1, x^2)$ and $x^- 
= x^0 - x^3$ are the ``spatial'' coordinates], by their Fourier
expansions in terms of creation and annihilation operators (see Appendixes
\ref{app:spinors} and \ref{app:xkapp}
 for details).  The canonical Hamiltonian derived this way 
leads to infinities in perturbative calculations of observables, due to 
summation 
over intermediate states with an infinite range of relative momenta that are 
reached by the local interactions from any state with only finite relative 
momenta. In order to regulate the Hamiltonian {\em en bloc}, including the 
binding 
mechanism, one has to introduce a cutoff $\De$ on the relative momenta
in the interaction terms. The cutoff requires counterterms that remove 
the cutoff-dependence from observables in the limit $\De\rightarrow\infty$. 
Thus, the Yukawa-theory Hamiltonian takes the form
\begin{equation}
\label{can:hd}
H^\Delta=H_0+H_Y^\Delta+H_+^\Delta+X^\Delta\;,
\end{equation}
where $H_0$ denotes the contribution of the first two  terms in Eq. 
(\ref{can:pminus}), $H_Y$ of the third term, and $H_+$ of the fourth one. 
$X^\Delta$ represents the counterterms. To be specific, the relative 
momenta (see Appendix \ref{app:xkapp} for definitions of $x$ and 
$\kappa^\perp$) of created or annihilated particles are limited by 
inserting factors $r_\De=\exp{(-\kappa^{\perp 2}/\De^2)} r_\delta(x)$ 
in the interactions terms, e.g. see Eq. (\ref{can:hy}). In the current 
study of Yukawa theory, the $x$-regulator factor $r_\delta(x)$ is  needed 
only for the mass-counterterm construction. In all other formulae, it can 
be replaced by $1$.  Since $r_\delta(x)$ does not appear in the final 
results, its form is left unspecified here (cf. \cite{Ggluon}).

The free part, $H_0$, is
\begin{eqnarray}
\label{can:h0}
 H_0&=&\sum_{i,\sigma} \int[p] \frac{p^{\perp 2}+m^2}{p^+} \left(
b^{(i)\dagger}_{p\sigma}b^{(i)}_{p\sigma}+
d^{(i)\dagger}_{p\sigma}d^{(i)}_{p\sigma} \right) \nonumber\\
&&+\int [p]\frac{p^{\perp 2}}{p^+}a^\dagger a\;.
\end{eqnarray}
There are three interaction terms. The one relevant here
consists of emission and absorption of a boson by the fermions, 
i.e.
\begin{widetext}
\begin{eqnarray}
\label{can:hy}
H_Y^\De&=&g\sum_{i,\sigma_1,\sigma_2}\int\left[p_1p_2q \right]
2(2\pi)^3 \delta^3\left( p_1-p_2-q\right)  \exp\left(-\kappa^{\perp
2}_{p_2,q}/\Delta^2\right)r_\delta(x_q/x_{p_1}) \times \nonumber\\
&&\left[a^\dagger _q b^{(i)\dagger}_{p_2\sigma_2}b^{(i)}_{p_1\sigma_1}
\bar u_{p_2\sigma_2}u_{p_1\sigma_1} -a^\dagger _q
d^{(i)\dagger}_{p_2\sigma_2}d^{(i)}_{p_1\sigma_1} \bar
v_{p_1\sigma_1}v_{p_2\sigma_2} +h.c.  \right] \; .
\end{eqnarray}
\end{widetext}
The other two are: those that create fermion-antifermion pairs 
from a boson and vice versa, and the instantaneous interaction
between fermions and bosons mediated by fermions. These two 
interactions are not written explicitly here since they do not 
contribute to  equations analyzed in the following sections. 
It is quite possible, and actually desired, that these additional 
interactions will improve the constituent picture when included
in the dynamics. However, it cannot happen in calculations of
the lower order than 4th, which is not known yet. 

\section{Approach 1: Bound states of two bare fermions}
\label{sec:A1}\label{sec:bar}

This Section reviews the renormalized Tamm-Dancoff procedure
for two-fermion bound states. One starts with a single fermion
eigenvalue problem, and then proceeds to the two-fermion case.

\subsection{One-fermion eigenstates}
\label{sub:bar-1f}

The one-fermion eigenvalue equation is first obtained by assuming 
that the coupling constant in the theory is infinitesimally
small and the dominant part of the eigenstate is provided
by a single bare-fermion Fock state. The quantum numbers of
the lowest-mass eigenstate correspond then by definition to 
one physical fermion associated with the fermion field
in the initial Lagrangian. When the coupling constant is 
made finite and grows, the eigenvalue equation is not soluble
exactly with currently known mathematical methods, and one
has to investigate results that follow from various attempts
to find approximate solutions. One such attempt is made by 
reducing a cut-off dynamics to the one-bare fermion Fock sector
for finite coupling constants, too. In that case, the projection 
operator in the operator $R$ (see Sec. \ref{sub:R} and Appendix
\ref{app:red})
has the form $\hat P = \sum_\sigma\int[p] b_{p,\sigma}^{(1)\dagger}
\ket{0}\bra{0} b_{p,\sigma}^{(1)}$. For finite cutoffs and
sufficiently small coupling constants $g$, one uses expansion 
in powers of $g$ to evaluate the corresponding $H_R$. Up to order 
$g^2$, this leads to an equation $H_R\ket{k}=P^-\ket{k}$, with
\begin{equation}
P^-=\frac{k^{\perp 2} +m^2_\Delta}{k^+}+X_{ff}=: \frac{k^{\perp 2}
+m^2_{f}}{k^+} \; ,
\label{mf}
\end{equation}
where $m^2_\Delta$ results from emission and re-absorption of 
bosons and $X_{ff}$ is contributed by the counterterm proportional
to $b\hc b$. Since $m^2_\Delta$ is a diverging function of $\Delta$, 
$X_{ff}$ has to be adjusted to remove that effect. Note that 
$m^2_f$ should not and does not depend on the fermion momentum 
components $k^+$ and $k^\perp$ in the light-front theory developed 
here. The expression for the eigenvalue $P^-$ is unique to light-front 
Hamiltonian dynamics with regulators preserving boost symmetries, 
so that for all momenta $k^\perp$ and $k^+$ there is one and 
the same value of $m_f$. Common approaches to field theories in 
standard time quantization lead to momentum-dependent ``mass 
terms'', i.e. the candidates for $m_f$ in the eigenvalue $E_f(\vec k) 
= (m_f^2 + |\vec k|^2)^{1/2}$ depend on $\vec k$ and boost symmetry 
is violated. It is also common in those approaches to consider 
only slowly moving fermions, especially at rest with respect to the 
frame distinguished by the definition of $H^\Delta$, and to identify 
the value of the function $m_f(\vec k)$ at $0$ with the desired 
fermion mass. Such approach is at best only approximate in the case 
of $\vec k \neq 0$ and creates conceptual difficulties in description
of bound-state constituents with large relative momenta. Therefore,
the standard quantization is replaced here by the light-front scheme.

The result (\ref{mf}) for $m^2_\Delta$ requires a counterterm of 
the form
\begin{eqnarray}
\label{cterm}
X_2&=&\sum_{i=1}^2\sum_\sigma\int[p] b_{p\sigma}^{(i)\dagger}
b_{p\sigma}^{(i)} \frac {1}{p^+} \times
\nonumber\\
&& \frac{g^2}{16\pi^2} \left[
\frac{\De^2}{2}\int dx \frac{1}{x}r_\delta^2
+4m^2\log\frac{\De^2}{m^2}+C \right] ,
\end{eqnarray}
where the constant $C$ is a finite part of dimension $m^2$. This condition 
removes $\De$-dependence from the physical fermion mass $m_f$ in the limit 
$\De\rightarrow\infty$.

\subsection{Two-fermion bound states}
\label{sub:bar-2f}

Reduction (Sec.\ref{sub:R}) to a two-bare fermion Fock sector employs 
(note that the two fermions are selected to be of different kinds)
\begin{equation}
\hat P
= \sum_{\sigma_1\sigma_2} \int[p_1p_2]
b_{1}^{(1)\dagger}b_{2}^{(2)\dagger}\ket{0}\bra{0}
b_{2}^{(2)}b_{1}^{(1)}\;,
\end{equation}
and leads to an equation for an eigenstate of $H_R$, which can be written as
\begin{equation}
\ket{P\sigma} = \int [p_1p_2] P^+ (2\pi)^3 \delta^3(P-p_1-p_2) 
\phi_\sigma
b_{p_1}^{(1)\dagger}b_{p_2}^{(2)\dagger}\ket{0}.
\end{equation}
The eigenvalue equation can be then written in terms of the two-body wave 
function $\phi_\sigma\equiv\phi_\sigma(\vec k)$
($\sigma$ denotes spin quantum numbers of both fermions 
$\sigma=\{\sigma_1,\sigma_2\}$, for definition of $k_3$ 
that forms $\vec k$ together with $k^\perp$,
see Appendix \ref{app:xkapp}). Namely,
\begin{widetext}
\begin{equation}
\label{schrobar}
\frac{\vec k^2}{m_f} \phi_\sigma(\vec k) + \sum_{\sigma'}
\frac{m_f}{(2 \pi)^3}  \int \frac{d^3k'}{ \sqrt{E_k E_{k'}}}
v_{OBE}(\sigma,\sigma',\vec k,\vec k')\phi_{\sigma'}(\vec k')=
\frac{M^2_{full}-4m_f^2}{4m_f} \phi_\sigma(\vec{k})\;,
\end{equation}
\end{widetext}
where $E_k=(\vec k^2+m^2)^{1/2}$, and the potential kernel
$v_{OBE}$ will be discussed below.

Note the mass $m_{f}$ in Eq. (\ref{schrobar}), i.e.
the physical fermion mass obtained from the earlier reduction to 
one bare fermion space, Eq. (\ref{mf}). Expanding the bare mass $m$, 
in both the integration measure factor $(E_1E_2)^{-1/2}$ and 
potential $v_{OBE}$, in a series of powers of $g$ around 
$m_f$, leads to an equation in which only physical mass $m_f$ enters. 
This way the bound-state dynamics for two bare fermions is related 
to a physical fermion mass parameter. This step makes a connection 
between the bare fermions in the two-body problem and a physical 
fermion obtained in the one-body reduction in the previous subsection.

To simplify farther discussion of the two-fermion eigenvalue equation, 
we denote the single-fermion mass-eigenvalue 
$m_f$ by $m$ (i.e. the subscript $f$ is dropped). The 
two-fermion bound-state mass $M_{full}$  can be rewritten as 
$M_{full}=2m-E_B$. When $E_B\ll m$, the  eigenvalue takes the form 
$ (M^2_{full}-4m^2)/4m =-E_B+ E_B^2/4m\approx -E_B$. Therefore, the eigenvalue 
on the right-hand side of Eq.(\ref{schrobar}) can be thought of as the 
binding energy $E_B$.

\begin{figure}[htbp]
\begin{center}
\includegraphics[width=6cm]{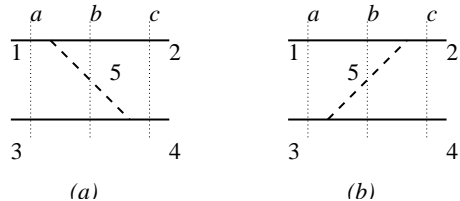}
\caption{Two kinds of terms in one boson exchange potentials. Following 
Ref.\cite{Ggluon}, the initial (rightmost), intermediate, and final 
(left-most), states are denoted by $c$, $b$, and $a$, respectively. 
For example, in diagram $(a)$, $p^+_{ba}=p^+_1,\;p^+_{bc}=p^+_4,
\;ba=M^2_{2+5}-m^2,\;bc=M^2_{3+5}-m^2,\; q^+=p^+_5$.  }
\label{fig:2diag}
\end{center}
\end{figure}

Since the regulator function $r_\De$ respects kinematical boost
invariance of the light-front scheme, this equation is  independent 
of the total momentum of two fermions. There is also no explicit 
$\De$-dependence in the matrix elements of the potential $v_{OBE}$ 
in the limit $\De\rightarrow\infty$:
\begin{eqnarray}
\label{OBE}
&\lim_{\De\rightarrow\infty}&
v_{OBE}(\sigma_1\sigma_2\sigma_3\sigma_4\vec k\vec k')=\nonumber\\
&&=-\frac{\pi\alpha}{2m^2} \frac{\bar u_1 u_2 \bar u_3 u_4}{q^+} \left(\frac
{p^+_{ba}}{ba}+\frac{p^+_{bc}}{bc}\right)_{Fig.\ref{fig:2diag}a}+
\nonumber\\ &&+\text{the same}|_{Fig.\ref{fig:2diag}b}\;,
\end{eqnarray}
where $\alpha=g^2/4\pi$.
The notation adopted here is taken from Ref. \cite{Ggluon} (see also caption of
Fig.~\ref{fig:2diag} here).
The spinor matrix elements are,
\begin{equation}
\bar u_{1} u_{2}=\frac{1}{\sqrt{x_1x_2}}\chi_1\hc[m(x_1+x_2)-
\sigma^3\sigma^\perp(x_1k_2^\perp - x_2k_1^\perp)]\chi_2,
\label{uu}
\end{equation}
where $\chi^\dagger=\left[1,0\right]$ or $\left[0,1\right]$, depending on
the fermion-spin projection on the $z$-axis.

The potential (\ref{OBE}) is a quite complicated, nonlocal function
of fermion momenta. But in the region with both momenta $k,k'\ll m$, 
it simplifies to the well-known Coulomb potential (cf. Appendix 
\ref{app:sch}),
\begin{equation}
\label{coulomb}
v_{Coulomb}=-4\pi\alpha \frac{1}{(\vec k -\vec k')^2}\;.
\end{equation}
Unfortunately, this heuristic result is not meaningful because the
region of large relative momenta of the fermions introduces important
corrections. From Eq. (\ref{OBE}) one can see that for one of the relative 
momenta ($k$ or $k'$) much bigger than the other, and than the fermion mass,
the spinor factors become proportional to the larger of the two momenta.
For example, if $k' \gg k,m$, one obtains $\bar u u \sim k'$, and two 
such factors compensate the denominator that grows
as ${k'}^2$ \cite{Glazek1980}. The potential becomes a function of 
$x_1$ and $x_2$, being a constant in the transverse momentum directions. 
A constant potential in the transverse momentum space is a two-dimensional 
$\delta$-function potential in configuration space. Such potentials with 
a negative coefficient lead to bound states of infinite negative energies
in the non-relativistic Schr\"odinger equation, and the light-front 
transverse dynamics is of this type. One could try to rely on 
the regulators $r_\De$ with finite $\De$ to remove the problem,  which 
would correspond to smearing of the $\delta$-potential in position space. 
The eigenvalues of the equation would then depend on $\De$. One could 
naturally try to make the coupling constant $g$ depend on $\Delta$. 
However, the interaction is specific to the Fock sector under consideration, 
it is much more complicated than a $\delta$-function itself, at least by the 
presence of the additional $x$-dependent factors,
and it is unlikely that a change of $g$ to a function of $\De$ can remove
the cutoff dependence from all eigenvalues. The 
procedure cannot be based on exact solutions in seeking simple remedies
for the cutoff-dependence. 

A perturbative search for suitable 
counterterms in the two-body sector was carried out up to second order 
in Ref. \cite{pinsky}. The problem of buildup of overlapping divergences 
in the two-body sector was in principle solved for low-mass eigenvalues 
in the limit $\Delta \rightarrow \infty$ in Ref. \cite{overlap}. However, 
that procedure was not able to deal with the small-energy denominators 
in a perturbative calculus for effective Hamiltonians. It is worth 
stressing that the cutoff-dependence problem 
appears not only when one sends $\De$ to infinity, but already for finite 
$\De$, since the eigenvalues have interpretation of $M^2$ and some of them
 become negative and lose physical 
interpretation when $\De$ and the coupling constant are large enough. 
This problem concerns lowest angular momentum states. Note also, that in 
some of the states, divergences may cancel out \cite{karmanov} due to 
interference of various spin amplitudes in the limit $\De \rightarrow \infty$.
Nevertheless, the two-body equation as it stands is not convergent in the 
large 
relative momentum domain, and the cutoff-dependence invalidates the 
non-relativistic approximation as a means for seeking a conceptually
satisfying solution of the divergence problem, especially for sizable 
coupling constants. 

A calculation described in the remaining part of this 
Section illustrates how the overlapping divergence problem arises in 
approach 1 in a quantitative way. An analogous calculation with effective 
fermions in Subsection \ref{sub:eff-2f}, will not produce such diverging 
results. This distinction is important
because the effective-particle approach allows then a self-consistent
farther development of the Fock-space expansion and inclusion of higher-order
terms in the Hamiltonians, which together may eventually lead to a well-founded
constituent picture.

Since the potential in the (non-relativistic) region of $k$ and $k'$ small 
in comparison to $m$ has the Coulombic form, one can ask with what accuracy 
Eq. (\ref{schrobar}) can be approximated by a Schr\"odinger equation with the 
Coulomb potential (given in Appendix \ref{app:sch}). 
The potential can be re-written in the form
\begin{equation}
v_{OBE}=v_{Coulomb}+\Delta v
\end{equation}
and corrections induced by $\Delta v$ estimated in perturbation theory. 
The Coulomb potential does not depend on spins of the interacting fermions.
Therefore, in $0$-th order of the bound-state perturbation theory, there are 
four degenerate states with the lowest mass $M$, and the same momentum-space 
wave functions: a triplet of spin-1 states, and a singlet of spin 0.

To estimate the first-order energy correction one has to find eigenvalues 
of the $4\times4$ matrix of matrix elements $\bra{\psi_{0i}}\Delta  
V\ket{\psi_{0j}}$, where $i$ and $j$ refer to the different spin 
configurations. The eigenstates of this matrix have spin structure: 
$(\uparrow\downarrow+\downarrow\uparrow)$, 
$(\uparrow\downarrow-\downarrow\uparrow)$,
$\uparrow\uparrow$ and $\downarrow\downarrow$. The  lowest mass
eigenstate is $(\uparrow\downarrow-\downarrow\uparrow)$.
The fist-order correction to the Coulomb energy for this state varies 
between numbers of the order of
$-4\;10^{-5}E_0$ for $\alpha=0.01$ to $-0.09 E_0$ for $\alpha=0.6$.
Note that $\alpha$ is also present in the wave function $\phi_0$ and the 
results are not connected by a straightforward multiplication by the ratio 
of the coupling constants, although both results are small, indeed. 
For $\alpha$ greater then $0.6$ the first-order correction would 
continue to be relatively small, but the second-order corrections (see
below) become unacceptably large, and this is why we do not discuss 
$\alpha$ significantly larger then $0.6$.

Unfortunately, in the second order a convergence problem in the domain of 
large relative momenta of fermions destroys self-consistency of the 
naive perturbative procedure around a non-relativistic approximation. To 
see this in a transparent way, one can make a number of 
simplifications and isolate the origin of corrections that grow with $\Delta$, 
without worrying about details of secondary importance. The point is that 
the Coulomb basis functions have quickly falling off tails in momentum space.
A tail is still small but greatly enhanced by first-order corrections,
and the second-order correction involves matrix elements that already diverge
with the cutoff $\Delta$.

To see the origin of the overlapping large-relative-momentum divergence in 
the second-order energy correction, one needs to analyze matrix elements of 
the type 
\begin{equation}
\Delta E^{(2)}=\bra{\phi_0}\De v \frac{1}{E_0-H_0-V_{Coulomb}}\De v
\ket{\phi_0} \; .
\label{DE2}
\end{equation}
Such elements involve integration over four relative momenta of fermions:  
the leftmost wave function argument denoted by $k_l$, the momentum of
states between the left $\Delta v$ and ${1}/({E_0-H_0-V_{Coulomb}})$,
denoted by $p_l$, the momentum between the operator 
${1}/({E_0-H_0-V_{Coulomb}})$ 
and right $\De v$, denoted by $p_r$, and the argument of the right wave 
function, denoted by $k_r$. The matrix element can be split into a sum of 
$2^4$ parts, with each part distinguished by saying how large is each of the 
four integrated momenta in comparison to the fermion mass $m$, smaller or 
larger.

Firstly, since the Coulomb wave functions strongly limit their
arguments, a part with $k_l$ and $k_r$ large is a very small 
contribution in comparison to the part with $k_l$ and $k_r$ small. 
Therefore, one looks for important contributions assuming 
that $k_l$ and $k_r$ lie within several widths of the Coulomb
wave functions. An {\em ad hoc} number used here is $4\alpha m$.

Secondly, there is no bound-state wave function limiting the intermediate 
momenta $p_l$ and $p_r$, and integrals over them extend up to the cutoff 
$\Delta \rightarrow \infty$. Eq. (\ref{uu}) shows that for large momenta 
the spin-flip part of the potential dominates other parts. This dominant part 
is selected here and denoted by $\De v_{\downarrow\uparrow}$,
both fermions have opposite spin orientation and both have their spins flipped 
in the interaction. For the purpose 
of estimating the order of magnitude of the large-momentum spin-flip 
contribution, $p_l$ and $p_r$ are considered larger than $m$ and 
$(E_0-H_0-V_{Coulomb})^{-1}$ is replaced by $-1/H_0$, neglecting terms that 
would vanish when $\alpha \rightarrow 0$. Then, the resolvent becomes diagonal
in momentum space and $p_l = p_r$ are commonly denoted by $k_2$. Details of 
how the cutoff $\Delta$ was initially introduced are not important for the 
order of magnitude estimate. Therefore, the cutoff function is slightly 
changed to simplify the integration. Namely, the 
initial $r_\De$ limits changes of invariant masses in each of the vertices,
see Fig.~\ref{fig:2diag}, producing a complex shape of the $k_2$-integration
boundary with details that depend on the small momenta $k_l$ and $k_r$, 
irrelevant to the divergence issue at hand. The main role of the cutoff in 
$\Delta E^{(2)}$ is to provide the upper limit on the range of integration 
over $k_2$. Therefore, one can estimate the size of the large-momentum range 
contribution by introducing a new cutoff $k_{max}$, equal to the maximum value 
that $k_2$ can take, and let $k_{max} \rightarrow \infty$ when 
$\De\rightarrow\infty$. The dependence on $k_{max}$ will indicate dependence
on $\De$. By the way, introducing a $k_{max}$ in $\Delta E^{(2)}$
is also a way one could introduce cutoffs if regularization were first 
imposed in the reduced eigenvalue equation (\ref{schrobar}) rather than in 
the initial QFT Hamiltonian, cf. \cite{karmanov}. Using the simplification of
$r_\De$ to $\theta(k_{max}-|\vec k_2|)$ here should not be understood as 
advocation of that procedure in a precise QFT calculation.

\begin{figure}[htbp]
\begin{center}
\includegraphics[width=6cm]{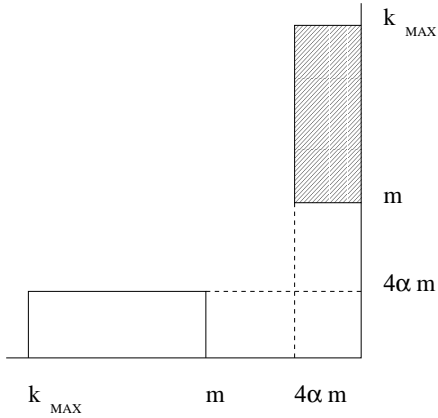}
\caption{The shaded area represents the range of integration over
$|\vec k_1|$ and $|\vec k_2|$ in 
Eq.(\ref{bareE2}) with
the potential matrix $\De v_{\downarrow\uparrow} 
(\vec k_1,\vec k_2)$. In the lower-right corner, both momenta are equal zero.}
\label{fig:rogi}
\end{center}
\end{figure}

\begin{figure}[htbp]
\begin{center}
\includegraphics[width=8.5cm]{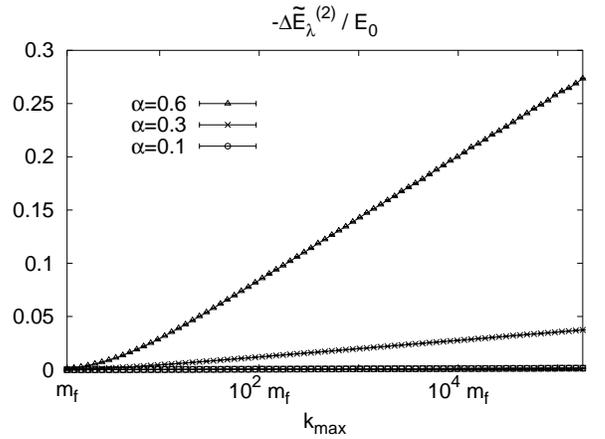}
\caption{Dependence of the most singular part of the 2nd-order
correction  (Eq. \ref{bareE2}) on the cutoff $k_{max}$. This
correction diverges logarithmically for large $k_{max}$
even for small coupling constants $\alpha$, though the 
matrix elements of the two-body Hamiltonian do not depend
on $k_{max}$ for momenta smaller than $k_{max}$.}
\label{fig:bezf}
\end{center}
\end{figure}

As a result of these steps, the large relative-momentum part of the 
second-order 
energy correction can be estimated from the following expression (the tilde 
indicates the simplifications made in $\Delta E^{(2)}$ as discussed above).
\begin{widetext}
\begin{equation}
\label{bareE2}
-\De \tilde E^{(2)} =m^2\int_0^{4\alpha m} \frac{d^3k_1}{\sqrt{E_1}}
\int_m^{k_{max}} \frac{d^3k_2}{E_2} \int_0^{4\alpha m}
\frac{d^3k_3}{\sqrt{E_3}} \phi_0(k_1)\De
v_{\uparrow\downarrow} (\vec k_1,\vec k_2)
\frac{1}{H_0}  \De v_{\uparrow\downarrow}(\vec
k_2,\vec k_3)\phi_0(k_3)
\end{equation}
\end{widetext}
The range of integration over $k_2$ in this expression, is shown in 
Fig.~\ref{fig:rogi}. 
Since the potential $\De v_{\uparrow\downarrow}$ 
approaches a constant for $k_2 \gg m$, one can expect a logarithmic 
dependence of $\De \tilde E^{(2)}$  on $k_{max}$,
\begin{equation}
\int_m^{k_{max}} \frac{d^3k_2}{E_2} \De
v_{\uparrow\downarrow}(\vec k_1,\vec k_2)
\frac{1}{H_0}\De  v_{\uparrow\downarrow}(\vec
k_2,\vec k_3) \sim
\log\frac{k_{max}}{m}\;.
\end{equation}
A numerical evaluation of the 12-dimensional integral produces 
an estimate of the actual size of the logarithmically diverging
correction. The results for different values of the coupling constant
are given in Figure \ref{fig:bezf}. The errorbars indicate the standard
deviation of a Monte Carlo routine used in the computation.
All other parts of the second order two-fermion bound-state mass 
correction (i.e. the parts with external momenta bigger than 
$4\alpha m$, or internal momenta smaller than $m$, or parts without 
change of the fermion spins) cannot compensate this divergence.
Note that the corrections can quickly reach the order of 10\% 
for coupling constants of the size expected in quark physics when the 
cutoffs are made larger than 100 quark masses. Even if one assumes
that $m \sim 1/3 $ GeV, momentum transfers on the order of 30 GeV
would be already far too large to fit under such cutoffs. The 
constituent wave-function picture would encounter serious conflicts 
with first principles of the underlying theory much earlier. If one
attempted an analysis starting from much smaller values of the quark
masses, closer to the Standard Model, the allowed cutoffs could
become extremely small and the procedure would be stuck with huge
corrections from the large-relative momentum region, out of 
control.
 
An analysis analogous to the approach 1, though more advanced with 
respect to the actual bound state dynamics (thanks to solving the singular 
eigenvalue equations numerically) was carried out in Ref. \cite{pinsky}. 
Those authors used the renormalization group ideas and calculated 
in perturbation theory two-body counterterms that could remove the 
cutoff dependence from the bound-state eigenvalues to large extent. 
Their procedure assumed that eventually the renormalization triangle 
with growing cutoffs and particle number could be obtained in a much 
larger project with many Fock components, but they 
did not go beyond their first step since the counterterms in higher 
orders (when the potential would be more complicated) would be much 
more difficult to find and understand. Also, if the counterterms are 
guessed sector by sector, a clear connection to the initial field 
theory is not directly visible and great determination would be 
required to pursue studies of such complex eigenvalue 
problems and their dependence on the cutoffs.  It is also worth stressing 
that for $\alpha$ small enough, there exist eigenstates of the
low-angular momentum two-body problem with finite eigenvalues $M^2$ 
even for $\De\rightarrow \infty$ \cite{karmanov}. This does not change
the fact that important eigenvalues in the bound-state equations for
bare constituents depend on the cutoffs when the latter are much larger 
than the fermion masses. It is hard to guess a self-consistent path out 
of the problem without systematic understanding of many relativistic 
QFT effects that currently appear to lie far beyond a few-body 
approximation.

Nevertheless, one could try to farther develop standard perturbative 
methods of constructing counterterms using Wilson's renormalization 
group idea \cite{Wold} in the approach 1. But in the presence of 
spin-induced divergences 
in the bound states of fermions, there exists a problem of degeneracy
of states that, most probably, cannot be resolved without use of the 
newer idea called the similarity renormalization group procedure \cite{sim1}.
This new method avoids the small energy denominator problem in degenerate 
perturbation theory. It can be used in evaluating effective Hamiltonians 
with small renormalization group parameters, and these parameters can
then play the role of cutoffs that are sufficiently small to define the 
effective constituent dynamics self-consistently. Since the similarity 
idea is also one of the key ingredients of the effective-particle approach 
used in the next Section, it is important to point out here in general 
terms how it differs from the standard procedure of integrating out  
high-energy degrees of freedom for bare particles. 

\begin{figure}[htbp]
\begin{center}
\includegraphics[width=8cm]{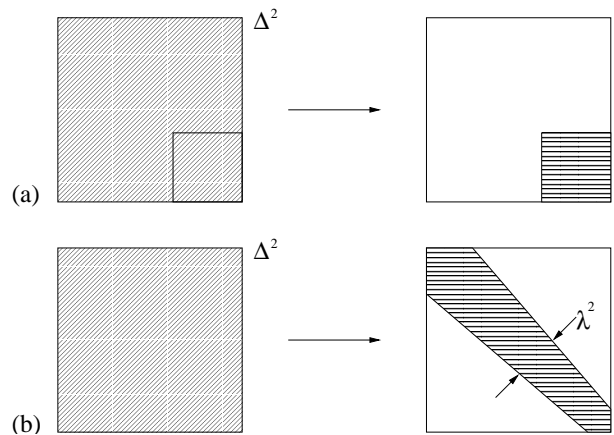}
\caption{(a) The standard renormalization procedure is based on 
reduction of a space of states. This leads to small energy 
denominators in perturbation theory.  (b) Similarity renormalization 
procedure is based on a rotation of basis and avoids small 
denominators in perturbative derivation of effective dynamics, 
because it integrates out only these energy changes that are larger 
than $\la$.}
\label{fig:wilsonvssim}
\end{center}
\end{figure}

In the presence of divergences, the standard requirement is that counterterms 
are chosen in such a way that there is no $\Delta$-dependence in a model 
Hamiltonian acting in a subspace of limited "kinetic" energy, 
see Fig.~\ref{fig:wilsonvssim}a.
The problem is that if there is no finite energy gap in the basis of
eigenstates of $H_0$ between the retained and integrated out states, then 
small energy differences between states in the retained subspace and 
outside of it appear in denominators in perturbative transformation $R$ 
that is repeatedly used in the renormalization group part of the whole 
calculation. The small denominators lead to large numbers multiplying terms 
which depend on $\Delta$, and intruder states can spoil the utility of that 
procedure completely. 
Therefore, to cancel the $\Delta$-dependence by the calculated counterterms
one would have to make incredibly precise calculations for very large 
$\Delta$.  In the similarity approach, the construction of counterterms and 
the transformation to small cutoff Hamiltonians are defined quite differently, 
see Fig.~\ref{fig:wilsonvssim}b. Instead of integrating states out, one brings 
the Hamiltonian matrix elements close to the diagonal in the sense that the 
new interactions change eigenvalues of $H_0$ only by less than an arbitrary 
but finite $\lambda$. The small energy denominators are avoided because 
one never eliminates transitions with energy changes smaller than $\lambda$, 
and so the denominators are always bigger than $\lambda$. When one comes
to the step of reduction $R$ of $H_\lambda$ to a few-body bound-state 
eigenvalue problem, one already has a well defined effective theory with 
a small width $\lambda$ and, in principle, no trace of the diverging 
cutoff parameter $\Delta$. 

One more advantage of the similarity approach is that one can also
use it to introduce the notion of effective particles \cite{Gacta}
and switch over to the eigenvalue problem written in terms of them, instead
of the bare ones.
This option allows us to perform practical calculations in QFT 
in a new way described in the next Section.

\section{Approach 2: Bound states of two effective fermions}
\label{sec:A2}\label{sec:eff}

This Section briefly reviews the RGEP for 
deriving Hamiltonians $H_\lambda$ for effective particles of size 
$\la^{-1}$ and then applies $H_\lambda$ in the Yukawa theory to
a bound state of two effective fermions. The presentation 
refers to some steps made in the previous Section, but in the context of
effective particles instead of the bare ones.
Key differences are pointed out, that lead to  
the two-effective fermion dynamics that converges in the region
of large relative momenta. 

\subsection{Renormalization Group for Effective Particles}
\label{sub:eff-sim}
The RGEP is defined through a unitary rotation for creation and annihilation 
operators  \cite{Gacta}
\begin{equation}
\label{bla}
b^{\dagger}_\la=U_\la b^\dagger U^\dagger_\la.
\end{equation}
Each operator (such as a Hamiltonian) can be expressed in terms of
both sets of the operators: $b$ or $b_\la$, and it has different matrix 
elements in the Fock-space basis built using the operators of each kind. 
The idea of the RGEP is 
to perform the rotation (\ref{bla}) in such a way, that the Hamiltonian 
expressed in terms 
of $b_\la$, called the ``effective Hamiltonian of width $\la$'', 
denoted $\Heff$, contains vertex form factors $f_\la$ of width 
$\la$ in all interaction terms. There are infinitely many interaction 
terms in $H_\la$ for all finite values of $\la$ and their strengths 
vary with $\lambda$ \cite{Ggluon}. The choice for $f_\la$ made here 
is,
\begin{equation}
f_\la=\exp
\left[-\frac{(M_{created}^2-M_{annihilated}^2)^2}{\la^4}\right]\;,
\end{equation}
where $M_{created}$ is a total free mass of all particles 
created by a given term in $\Heff$, and $M_{annihilated}$ 
is a total free mass of particles annihilated by the term.

If the unitary transformation $U_\la$ were known exactly, there would 
be no $\la$ dependence in the spectrum of $\Heff$.  But when 
$U_\la$ (and $\Heff$) are calculated in perturbation theory, 
the approximation leads to some residual 
$\la$-dependence of theoretical predictions for observables. The sensitivity
of results to variation of $\la$, provides a simplest test for how large 
errors one makes in the perturbative expansion for $\Heff$, on top of
the error margin resulting from approximations used to solve the
Schr\"odinger equation with $\Heff$. On the one hand, one tries to
get down to as small $\lambda$ as possible so that the non-perturbative 
diagonalization will require smallest possible range of energy scales 
to handle explicitly, using a computer. On the other hand, one expects 
errors due to use of perturbation theory in evaluating $\Heff$
to grow with reducing $\lambda$, and $\lambda$ should not be too small. 
The reason is that if $\lambda \rightarrow 0$, the Hamiltonian becomes  
almost diagonal, which is equivalent to solving the non-perturbative 
dynamics of bound states, and a perturbative calculus for $\Heff$ 
must fail at some point before $\lambda$ becomes equal to the scale 
of the non-perturbative phenomena.

The same rotation $U_\lambda$ provides also means for constructing 
counterterms in $H^\Delta = H_{\la = \infty}$. A Hamiltonian $\Heff$
with a finite $\la$ has a band diagonal matrix in the effective basis,
and each effective Fock basis 
state is directly coupled only to a limited set of other effective 
states that have energies within the range of $\la$. Therefore,
the effective theory splits into a chain of theories that couple only to 
near neighbors, without jumping up to arbitrarily large scales such
as $\De$ in the approach 1. Consequently, if counterterms introduced 
in $H^\De$ cause that no $\De$-dependence appears in the matrix
elements of $\Heff$ in the effective basis states, there can be 
no $\De$-dependence generated in observables calculated in perturbation 
theory to any finite order when $\De/\la$ tends to infinity. This will
be seen in detail at the end of this Section, but it can be
observed already here that the bare-particle approach described in 
the previous Section did not have this property and $\De$-dependence
could show up in the bound-state dynamics even though matrix elements 
of the Hamiltonian reduced to the two-body sector did not depend
on $\De$ for all finite relative momenta.

\subsubsection{Effective Hamiltonian - 0th and 1st order}

The only change in the 0th-order Hamiltonian (free part, order $g^0$) is that
the bare operators such as $b^\dagger b$ are replaced by $b^\dagger_\la
b_\la$. In order $g^1$, the effective Hamiltonian has the form, 
\begin{widetext}
\begin{eqnarray}
\label{eff:hy}
H_\lambda^{\De(1)}&=&g\sum_{i=1}^2
\sum_{\sigma_1,\sigma_2}\int\left[p_1p_2q \right] 2(2\pi)^3
\delta^3\left( p_1-p_2-q\right)  \exp\left(-\kappa^{\perp
2}_{p_2,q}/\Delta^2\right)r_\delta(x_q/x_{p_1})
\exp\left[-\frac{\left(M_{p_2q}^2-m^2\right)^2}{\la^4}\right]
\nonumber\\ && \times\left[a^\dagger _{q\la}
b^{(i)\dagger}_{p_2\sigma_2\la}b^{(i)}_{p_1\sigma_1\la} \bar
u_{p_2\sigma_2}u_{p_1\sigma_1} -a^\dagger _{q\la}
d^{(i)\dagger}_{p_2\sigma_2\la}d^{(i)}_{p_1\sigma_1\la} \bar
v_{p_1\sigma_1}v_{p_2\sigma_2} +h.c. \right] \;.
\end{eqnarray}
\end{widetext}
Note that expressing $b$'s by $b_\la$'s induced the form factor
 $f_\la$ in $\Heff$.
This form factor causes that the regularization factor that depends on $\De$
is equivalent to 1 when $\De/\la \rightarrow \infty$. 

\subsubsection{Effective Hamiltonian - 2nd order: mass term}
One can calculate the term in $\Heff$ of order $g^2$ that contains $b_\la\hc
b_\la$ and see that it contains a mass-squared-like term with a divergent 
$\De$-dependence. Therefore, one has to add a counterterm to the initial 
Hamiltonian that has exactly the same form (\ref{cterm}) as in the approach 1. 
After including this counterterm, the form of the effective mass term in 
$\Heff$ is (in the limit $\Delta\rightarrow\infty$)
\begin{equation}
\label{eff:hbb}
H_{\lambda\delta m}=\int[p]b_\la\hc b_\la \frac{\delta m^2_\la}{p^+} \; ,
\end{equation}
where
\begin{eqnarray}
\label{eff:msq}
\delta m^2_\la&=&\frac{g^2}{16\pi^2}\int_{m^2}^\infty dz \frac{1}{2}
\left(1+\frac{6m^2}{z}+\frac{m^4}{z^2}\right)
\times\nonumber\\&&
\exp\left[-\frac{2(z-m^2)^2}{\la^4}\right] + const.
\end{eqnarray}
Note that the renormalization is carried out now at the level of full theory
in the whole Fock space, not after reduction to a specific Fock sector. 
Therefore, for example, there are no sector-dependent mass counterterms.
Since the regulators did not violate any kinematical light-front 
symmetries, the calculated mass term does not depend on particle momentum 
(i.e. the relativistic form of the dispersion relation does not change, 
there is only a change of the value of the effective fermion mass).

\subsubsection{Effective Hamiltonian - 2nd order: potential term}
Second-order terms in $\Heff$ that contain two creation and two
annihilation  operators for effective fermions do not contain 
any dependence on $\De$ when $\De\rightarrow\infty$, and no 
counterterms are needed of such form. Therefore, the complete
answer for these terms is 
\begin{widetext}
\begin{eqnarray}
H_{\la b\hc b\hc b b}^{\Delta(2)}&=&
\sum_{\sigma_1\sigma_2\sigma_3\sigma_4}
\int[p_1p_2p_3p_4]b^{(1)\dagger}_{p_1\la}b^{(2)\dagger}_{p_3\la}
b^{(2)}_{p_4\la}
b^{(1)}_{p_2\la}\; 2(2\pi)^3\delta^3(p_1+p_3-p_2-p_4)
\nonumber\\
&&\times
v^{(2)}_{\lambda}(p_1,p_2,p_3,p_4,\sigma_1,\sigma_2,\sigma_3,\sigma_4)\;,
\label{Hbbbb}
\end{eqnarray}
where
\begin{equation}
\label{vla}
v^{(2)}_{\lambda}(p_1,p_2,p_3,p_4,\sigma_1,\sigma_2,\sigma_3,\sigma_4)=
-g^2 \bar u_1 u_2 \bar u_3 u_4 \frac{1}{q^+}f_{ac} {\cal F}_2(a,b,c)|_
{Fig.\ref{fig:2diag}a} + \text{the same}|_{Fig.\ref{fig:2diag}b} \; ,
\end{equation}
\end{widetext}
and ${\cal F}_2(a,b,c)=[(x ba +(1-y)bc)/(ba^2+bc^2)](f_{ba}f_{bc}-1)
$, adopting conventions from Ref. \cite{Ggluon}
with $x=p_1^+/(p_1^++p_3^+)$ and $y=p_2^+/(p_2^++p_4^+)$. 
Despite that Fig.~\ref{fig:2diag} is referred to in order to define notation,
just like in approach 1, the potential $v_\la$ is 
quite different from the OBE potential of Eq. (\ref{OBE}).
For example, note the different kind of denominators and the presence
of the key formfactors $f_{ac}$.

\subsubsection{Effective Hamiltonian - 2nd order: other terms}

In the second-order effective interaction, there are also other terms,
similar  to the terms shown above, or describing interactions with explicit
participation of bosons. There are also terms creating or annihilating two
additional particles. None of these terms contribute to the second-order 
effective equation that will be obtained in the following Section in the 
case of bound states of two effective fermions. In general, the 
RGEP allows one to 
construct both the transformation $U_\la$ and Hamiltonian $H_\la$ in 
perturbation theory to an arbitrary order in $g$. Calculating higher-order 
corrections is ultimately the only way for finding out how large corrections 
they produce. The approach 2 is limited here
to terms of the second order, because it was the case in the approach 1.

\subsection{Solving eigenvalue problem with $\Heff$}

In the case of bound states of two effective fermions, the reduction
procedure is based on the same rules as in the approach 1, except that
the effective particles interact with vertex form factors of width
$\la$ and the large relative-momentum convergence is improved. Also
the change of particle number is severely limited in strength, since 
massive particles cease to be produced when $\la$ is lowered below their mass,
and the emission of massless particles changes energies by amounts linear
in the exchanged momentum. The changes of order $k$ are larger than 
the changes of order $k \times k/m$ in the fermions' energies when $\la$
is smaller than $m$. The departure point of the process of solving
the bound state dynamics is the eigenvalue equation for the single 
fermion states.

\subsubsection{Reduction to one effective fermion subspace}
\label{sub:eff-1f}
This step produces an equation $H_R\ket{k}=P^-\ket{k}$, where
\begin{equation}
P^-=\frac{k^{\perp 2} +m^2_{f}}{k^+}\;,
\end{equation}
and $m^2_{f}$ is the physical fermion mass, by definition of the same 
value as in the approach 1. It comes out independent of $\De$ by the 
virtue of adjusting once and for all the mass squared counterterm in 
$H^\De$. The same adjustment involves fixing the free finite constant 
in Eq. (\ref{eff:msq}) so that for some value of $\la = \la_0$ the 
physical fermion mass eigenvalue $m_{f}$ equals to the experimentally
found number. 
The interesting point is that the same eigenvalue is subsequently
automatically obtained for all values of $\la$ and the 
physical dispersion relation satisfies all requirements of special 
relativity. This is the  simplest manifestation of the general rule
that physical results should be independent of $\la$, as it is only a
parameter of a unitary rotation of the basis.  

\subsubsection{Reduction to two-effective fermions}
\label{sub:eff-2f}

Using transformation $R$ to reduce $\Heff$ to the two-effective particle 
subspace without restrictions on the relative momenta, one obtains a quantum
mechanical interaction that can change the invariant mass of the two
particles by a certain $\Lambda$ if, and only if the interaction 
acts more than $\Lambda/\lambda$ times. Thus, the approach 2 produces an 
effective Hamiltonian which is free from the overlapping divergence problem 
discussed in Ref.\cite{overlap}, 
and in the previous Section in approach 1. However, in order to make a
connection with the non-relativistic two-particle Schr\"odinger quantum 
mechanics, which was not systematically available in the approach 1, one needs 
now to limit the relative momenta in the effective two-particle Fock sector 
to $k < z$, where $z$ is a new parameter required for defining the new 
operation $R$ that enables one to define the procedure of introducing
the non-relativistic limit.

\begin{figure}[htbp]
\begin{center}
\includegraphics[width=8cm]{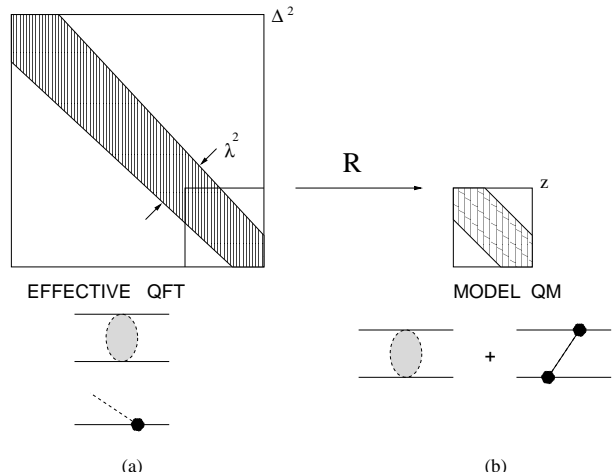}
\caption{Reduction of effective QFT, (a), to relativistic quantum mechanics 
in a model subspace, (b), is possible in the approach 2 thanks to the form 
factors in the effective interaction vertices.}
\label{fig:model}
\end{center}
\end{figure}

Therefore, a new transformation $R$ is now defined to lead to a model 
Hamiltonian $H_R$ that acts only in the subspace of the two-effective 
particles Fock sector with limited invariant masses (Fig.~\ref{fig:model}). 
Thus, not only the number of effective particles is limited, but also the 
range of their relative momenta. It is required that $H_R$ has the same 
spectrum of low lying energy levels as $\Heff$ has in the whole space. 
This step is no longer related in any way with infinite renormalization 
problem as in the approach 1. The existence of such reduction is plausible 
only because $\Heff$ has a small width $\lambda$ and in subsequent orders 
of perturbation theory in $H_{I \la}$, 
corrections to the effective potential result from the coupling to an
additional region of relative momenta, which is always limited by $\la$ in
every new order.

The projection operator used here is
\begin{equation}
\hat P
= \sum_{\sigma_1\sigma_2} \int[p_1p_2]
b_{1\lambda}^{(1)\dagger}b_{2\lambda}^{(2)\dagger}\ket{0}\bra{0}
b_{2\lambda}^{(2)}b_{1\lambda}^{(1)}\theta\left(z-|\vec k|\right)\;,
\end{equation}
where $\vec k$ is the relative momentum of effective particles of momenta 
$p_1$ and  $p_2$. Although introduction of $z$ is useful from the conceptual
point of view, the formfactors $f_\lambda$ imply that $z$ is not important 
in practice if only the lowest order
(i.e. $g^2$) model Hamiltonian is calculated (see Fig.~\ref{fig:model}). 
I would, however, affect the model Hamiltonian in order $g^4$ through terms 
such as the last two terms in Eq. (\ref{app:hr}).

The effective Schr\"odinger equation has then the form of Eq. (\ref{schrobar})
with $v_{OBE}$ replaced by a new potential, denoted $v_{R\la}$, which is a 
sum of two terms (Fig.~\ref{fig:model}b).  The first term is the projection 
of $H_{\la b\hc b\hc b b}^{\Delta(2)}$, cf. Eq. (\ref{Hbbbb}), on the two-body
space restricted by $z$. The second term comes from the one-effective boson 
exchange (OEBE) and has a form similar to (\ref{OBE}),
\begin{widetext}
\begin{equation}
\label{OEBE}
v_{OEBE\lambda}(\sigma_1\sigma_2\sigma_3\sigma_4\vec k\vec k')=
-\frac{\pi\alpha}{2m^2}{\bmnie {f_\la}}\bar u_1 u_2 {\bmnie {f_\la}}\bar u_3
u_4 \frac{1}{q^+}\left(\frac {p^+_{ba}}{ba}+\frac{p^+_{bc}}{bc}\right)
_{Fig.\ref{fig:2diag}a}+ \text{the same}|_{Fig.\ref{fig:2diag}b} \; ,
\end{equation}
\end{widetext}
except for the form factors $f_\la$ in vertices and the overall limitation
of the momenta by $z$ [not indicated explicitly in Eq (\ref{OEBE})]. 
Each of these two terms (i.e. projection of 
$H_{\la b\hc b\hc b b}^{\Delta(2)}$ and $v_{OEBE}$) behave 
for $k,k'\ll m$ like the Coulomb potential (\ref{coulomb}) with form factors 
$f_\la$ that limit changes of the fermion kinetic energies.
\begin{figure}[htbp]
\begin{center}
\includegraphics[width=8.5cm]{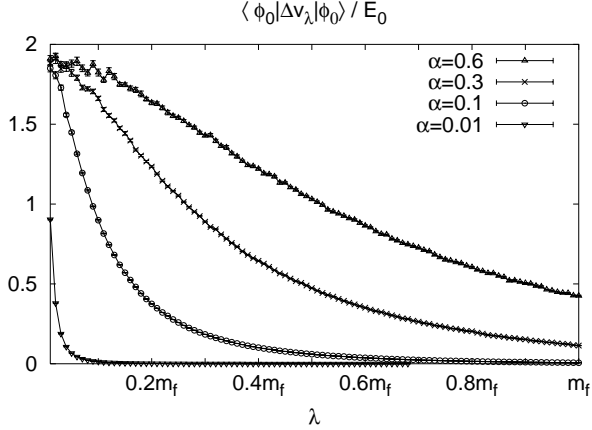}
\caption{First-order correction to the ground state  binding energy
as a function of $\la$ for $z=\infty$, which shows the magnitude of corrections
 to the well
known Schr\"odinger equation with a Coulomb potential
expected in the approach 2 in QFT.}
\label{fig:e1odlambda}
\end{center}
\end{figure}
One can approximate the Schr\"odinger equation with this QFT potential 
by the equation with a Coulomb potential plus a correction, and one
can estimate the size of the correction using bound-state perturbation 
theory. For this purpose, the difference between potentials $v_{R\la}$ and 
$v_{Coulomb}$ is denoted by  $\De v_\la$.  The first-order correction, 
$\De E^{(1)}_\la=\bra{\phi_0}\Delta v_\lambda\ket{\phi_0}$,
is a  function of the parameters $\la$ and $z$. Numerical calculation
confirms that for $z>\la$ there is no noticeable $z$-dependence 
of this  matrix element.
Fig.~\ref{fig:e1odlambda} shows how the matrix element depends on $\la$ 
for $z=\infty$. As expected in Section  \ref{sub:eff-sim} for small $\la$, 
there is some $\la$-dependence in the result. It emerges because
at too small $\la$s the similarity factors $f_\la$ start to limit 
the Hamiltonian in the  momentum region that is important for the 
bound state formation, and the derivation of $\Heff$ cannot be carried
out precisely using the perturbative renormalization group procedure
down to so small $\lambda$s.
When $\la$ and  $z$ are large enough, the correction $\De
E^{(1)}_\la$ approaches a fixed finite value, that depends on
$\alpha$. This happens
because the wave function $\phi_0$ has a width $a=\alpha \mu$  and
limits the integration over both momenta in the matrix element
$\bra{\phi_0}\Delta v_\lambda\ket{\phi_0}$. Thus, as seen already in
Section \ref{sub:bar-2f}, the first-order correction is small for
small coupling constants due to the fast fall-off of the Coulomb wave 
function at large momenta, independently of the details 
of $\De v$ that one obtains in the approaches 1 or 2. The 
correction is small even for a divergent potential such as 
a $\delta$-function.

Therefore, one needs to look at the second order of the bound-state
 perturbation theory
to check the self-consistency of the effective particle picture and
to compare it to the approach 1. To see that the effective 
theory does not exhibit the consistency problems the approach 1 
exhibited in Fig.~\ref{fig:bezf}, one 
can closely follow here the derivation of Eq. (\ref{bareE2}), but with the 
OBE potential $v_{OBE}$ replaced by $v_{R\la}$.  Again, one can ask 
whether there is a logarithmically divergent dependence on $k_{max}$.

\begin{figure}[htbp]
\begin{center}
\includegraphics[width=8.5cm]{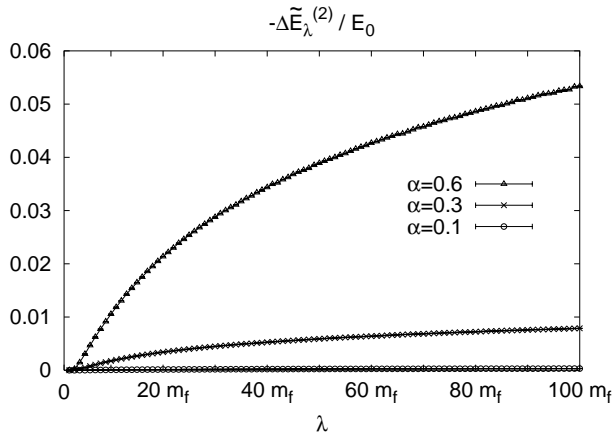}
\caption{Dependence of the large-relative momentum contribution to 
the second order bound-state mass correction on $\la$, for the cutoff 
$\De$ sent to $\infty$. When one works with effective fermions,
cutoffs can be sent to $\infty$ for any given value of $\lambda$,
which is not available in the approach 1 for bare fermions, as shown 
in Fig.{\ref{fig:bezf}}.}
\label{fig:e2odlambda}
\end{center}
\end{figure}

It turns out that for finite values of $\la$ there is no such
divergent dependence.  One can safely take the limit 
$k_{max}\rightarrow\infty$, since $\la$ itself already cuts off 
sums over intermediate states in the correction.
\begin{widetext}
\begin{equation}
\label{effE2}
-\De \tilde E_{\lambda}^{(2)} =m^2\int_0^{4 \alpha m}
 \frac{d^3k_1}{\sqrt{E_1}} \int_m^{\infty}
 \frac{d^3k_2}{E_2} \int_0^{4\alpha m}
 \frac{d^3k_3}{\sqrt{ E_3}} \phi_0(k_1)\De
 v_{\uparrow\downarrow\lambda} (\vec k_1,\vec k_2)
 \frac{1}{H_0}  \De
 v_{\uparrow\downarrow\lambda}(\vec k_2,\vec k_3)
 \phi_0(k_3)
\end{equation}
\end{widetext}
Here, 
$\De v_{\uparrow\downarrow\la}$ is defined similarly to
$\De v_{\uparrow\downarrow}$, but with $v_{OBE}$ replaced by
$v_{R\la}$. Numerical results for this matrix element for different 
values of $\la$ (and for the cutoffs $k_{max} \sim \De \rightarrow \infty$), 
are shown in Fig.~\ref{fig:e2odlambda}.

First of all, the results in Fig.~\ref{fig:e2odlambda} can be considered 
a good approximation to the whole 2nd-order correction only for $\la \gg m$ 
(i.e. in the right part of the figure). If $\la$ is comparable to $m$, 
the similarity factors $f_\la$ limit the potential $v_{R\la}$ and the 
high-low and low-high corners of the potential matrix 
(Fig.~\ref{fig:rogi}) are practically eliminated. The correction coming from 
the large momentum region selected in the integration in Eq. (\ref{effE2}) is 
therefore
also reduced and the other of the $2^4$ parts of the whole correction can 
contribute in more significant ways than they do for large $\la$. Hence,
for small $\la$, the results given in Fig.~\ref{fig:e2odlambda} are not 
necessarily a good approximation of the whole second-order energy-correction.

Secondly, in practical work, one needs to lower $\la$ as far down as 
possible, possibly below $m$. Thus, Fig.~\ref{fig:e2odlambda} provides
only evidence for the self-consistency of the effective 
fermion dynamics in which the convergence in the large-relative momentum 
region is secured by the presence of $\la$, and the original QFT cutoffs 
can be safely sent to infinity. 

Thirdly, despite the reservations made above, one can expect that the 
exact 2nd-order correction from the large momentum region 
to the bound-state mass is of the same order of magnitude 
as the part given in Fig.~\ref{fig:e2odlambda}. It is clear then that the 
2nd-order contribution from the large-relative momentum range 
is at least one order of magnitude smaller 
than the 1st-order result, presented in Fig.~\ref{fig:e1odlambda}.  Therefore, 
it is the small and moderate momentum region that decides about the size of 
variations of observables versus $\la$. In view of these three comments the 
following farther remarks can be made.

It is visible in Fig.~\ref{fig:e1odlambda} in the case of $\alpha=0.01$ that 
the bound state energy is a very slowly varying function of $\la$ when the 
latter grows above $0.2m_f$. One could call this value of 
$\alpha$ a non-relativistic coupling, because all the important dynamics 
happens among virtual effective particles with momenta much smaller 
than the fermion mass. Consequently, if the form factor $f_\la$ is wider 
than $0.2 m_f$, its
presence is hardly seen in the final result. Nevertheless, its presence 
remains to be essential for keeping the 2nd-order correction finite and 
making the large-momentum region of QFT a small correction 
(Fig.~\ref{fig:e2odlambda}) to the Schr\"odinger picture when 
$\De\rightarrow\infty$, instead of diverging and invalidating that picture, 
as it was happening in the approach 1 (Fig.~\ref{fig:bezf}). 
The system of effective particles is then shown 
to be self-consistently non-relativistic and the Schr\"odinger equation with  
Coulomb potential is a good approximation of QFT. Qualitatively, this is the 
situation encountered in QED with $\alpha = 1/137$.

However, as $\alpha$ increases, one has to make $\la$ larger to achieve a fair
$\la$-independence of the total binding energy, by reducing the size of the
$\la$-dependent corrections. Fig.~\ref{fig:e1odlambda} shows that the 
first-order corrections can be quite large and they depend on $\la$.
For $\alpha=0.1$, the correction has 
a non-negligible value ($\De E^{(1)}=0.06 E_0$) even for $\la=m_f$. 
For $\alpha=0.3$ and $\la=m_f$, $\De E^{(1)}=0.12 E_0$.
And if one wanted to make a reduction of the theory to the two-fermion 
state for $\alpha=0.6$, one would have to include momenta much larger 
than the fermion mass (for $\la=m_f$, $\De E^{(1)}=0.43 E_0$, and it 
still strongly depends on $\la$). In this case, the QFT effective two-body 
equation is no longer close to the non-relativistic Schr\"odinger 
equation with the Coulomb potential. Also, strong $\la$-dependence
suggests that considerable errors can come from the limitation to only 
second-order perturbative 
calculation of the similarity rotation. Besides, one should investigate 
whether perturbation theory for inclusion of higher effective Fock 
sectors is acceptable when solving the Schr\"odinger equation.

Fortunately, as one can see from Fig.~\ref{fig:e2odlambda}, at the same 
time the second-order corrections coming from large 
momenta remain small, even for $\alpha = 0.6$.
This is an important result, because it suggests that
preparing numerical procedures 
for solving effective particle dynamics with a few Fock sectors and seeking
better starting points for the bound-state perturbation theory than the pure 
Coulomb picture, is a good idea to try. Namely, if the region of large 
relative momenta were significant (and not manageable in perturbation theory, 
contributing too much in the second order), there would be no reason 
for why limiting 
the number of Fock sectors should be a legitimate approximation. The 
uncertainty principle would rather suggest that if momenta larger than
$m_f$ are important then multiparticle states are also important. 
Our result says, that thanks to $f_\la$ one can expect the 
effective Fock space expansion to be a legitimate strategy in the approach 2, 
although it requires checking.

Finally, the difference between the initial cutoff $\De$ and the
parameter $\la$ is important for a self-consistent interpretation
of the theory. The cutoff $\De$ is an artificial parameter that
chops off the high-energy part of the initial QFT Hamiltonian. One
wants to send $\De$ to infinity and one chooses counterterms in
$H^\De$ in such a way that observables are independent of $\De$ in
the limit $\De \rightarrow \infty$. The same huge cutoff appears
in the Tamm-Dancoff procedure with bare particles. There is then
no alternative in the approach 1 to sending $\De$ to infinity
in the reduction to two-bare fermions Fock sector. This is odd 
because there is no good reason for only a few bare particles to 
be relevant in the bound-state dynamics with the huge momentum cutoff. 
Indeed, new overlapping divergences are obtained this way. 
Without tools to handle the problem, no self-consistent
relativistic treatment of bound states of fermionic 
constituents in QFT can be reached that way (see also 
Section \ref{sub:bar-1f}).

In the conceptually different approach 2, $\De$ is absent in the effective 
particle dynamics, and the new finite width parameter $\la$ is introduced.
One adjusts $\la$ to contain the actual physics of the theory 
in most economic way, and the key measure of success in this respect,
besides verifying symmetries, is provided by the size of changes 
in the physical results when $\la$ is changed. Comparing the first
and second-order results from Figs. \ref{fig:e1odlambda} and 
\ref{fig:e2odlambda}, one can see that there is a whole range of 
$\la$s and coupling constants of considerable interest where stability 
of results versus $\la$ can be expected in more advanced calculations 
than used in the present pilot study. The renormalization group flow of 
$\Heff$ produces important corrections to the well known 
non-relativistic picture, visible in the significant first-order 
corrections, while the large-relative momentum region is 
self-consistently protected from divergences.
This way, a relativistic treatment of bound states
of fermions in the Fock space has a chance to be 
developed \cite{poincare}.

\section{Conclusions}
\label{sec:con}

One can attempt to calculate wave functions of bound states of  fermions 
in the Fock space representation of  the Yukawa QFT in various ways that 
are described and compared  in the previous Sections  on the example of 
bound states of two fermions. Two basic concepts are distinguished. One 
approach starts from the sector of two bare fermions (approach 1), and 
another one from two effective fermions (approach 2). 

Approach 1 leads to overlapping divergences in the light-front Hamiltonian 
dynamics and lacks self-consistency in handling the large-relative momentum 
region when one attempts to send the bare cutoff to infinity without including 
infinitely many bare particles. There exists  an option  of removing this 
defect through sector dependent counterterms, but the required construction of 
the full renormalization group triangle with growing numbers of bare particles 
appears overwhelmingly complex and certainly not completely understood. The 
basic ultraviolet problem comes from short distances in the transverse 
directions and no practical tool exists yet for handling huge numbers of bare 
particles with precision required by rotational, parity, and other symmetries 
of the initial Lagrangian.

Approach 2 is free from the difficulties with large-relative momentum 
convergence of the approach 1. The decisive convergence factor is introduced
through solving  renormalization group equations for effective particles. The
solution includes form factors of width $\la$ in the interaction vertices and 
the form factors suppress the large momentum domain. This is verified in
numerical estimates described in previous Sections. The well-known 
one-boson-exchange (OBE) potentials that are deduced from the on-shell 
$S$-matrix elements, are replaced by new one-effective-boson-exchange (OEBE) 
potentials, and additional interactions, that are derived in the Hamiltonian 
form independently of the corresponding $S$-matrix. Moreover, one can take
advantage of subsequent $S$-matrix calculation in choosing free finite
parts of the counterterms. Then the renormalized 
Hamiltonians in the QFT Fock space, include relativistic and off-energy-shell 
corrections in accordance with first principles of quantum mechanics, special 
relativity, and renormalization group. However, these principles are satisfied 
only approximately, since the solutions for $H_\la$ are found only order by 
order in perturbation theory, and diagonalization of the reduced Hamiltonians 
$H_R$ is carried out numerically. The approximate character is not a weakness,
however, because corrections can be systematically investigated and reduced.

The accuracy of the approximate treatment can be estimated by inspecting 
variation of results when one changes the renormalization group parameter 
$\la$. Exact theory would exhibit no dependence on $\la$. In the approximate 
treatment, variations depend on the size of the coupling constants.

For $\alpha$ smaller than about 0.3, there is a wide range of $\la$s in the 
Yukawa theory with massless bosons, in which the results for the 
two-fermion bound state mass are stable and do not differ much from results 
obtained from the nonrelativistic  Schr\"odinger equation with Coulomb 
potential. These values of $\alpha$ can be called non-relativistic. For 
$\alpha$ larger than about 0.3, one has to allow $\la$ to grow up to the size
on  the order of  $m_f$,  to  achieve $\lambda$-independence of the 
corrections to the bound-state mass. This means that relativistic momenta 
do matter and the non-relativistic Schr\"odinger equation is not  a good 
approximation of the effective dynamics.  Still, the effective
theory is well contained in the range determined by $\la$ and one can look 
for  solutions of the eigenvalue problem without making non-relativistic 
approximation. For the first-order bound-state perturbation-theory 
corrections to the non-relativistic approximation are quite considerable, 
while the large-relative momentum region in second-order corrections 
contributes merely about 10\%. Being limited by $\la$, it does not introduce
any diverging contributions and one does not and should not attempt sending
$\la$ to infinity, in contrast to the cutoff  $\Delta \rightarrow \infty$ in 
the approach 1.

All effective-particle calculations are carried out here in the light-front 
form of 
Hamiltonian  dynamics, preserving its kinematical covariance under boosts.
This form enables one to separate the relativistic motion of bound states 
from their internal constituent dynamics. Thanks to this separation, one can 
reduce the description of the binding mechanism to the Schr\"odinger 
equation for internal 
motion of the constituents without choosing a specific frame of
reference. All eigenvalue equations derived here are independent of  the total 
momentum of the states they describe. Thus, the boost symmetry  allows 
one to understand  moving bound  states in arbitrary motion as soon as one 
understands them at rest, which can help in understanding their structure in 
the infinite momentum frame, the opportunity  not available in the standard 
form of Hamiltonian dynamics.  In fact, the effective particle dynamics is in 
principle well-prepared to tackle Poincar\'e symmetry issues in the Fock 
space including both the kinematical and dynamical transformations 
\cite{poincare}.

When  the effective light-front theory is defined in a perturbative way,
the vacuum structure remains simple, i.e.  $b_\la |0\rangle  
= d_\la |0\rangle = a_\la |0\rangle = 0$.  In the calculations reported 
in this paper, in Yukawa theory, no vacuum structure appeared. In the case of 
theories 
where spontaneous symmetry breaking  may occur, one has to include new terms 
in the initial Hamiltonian that contain the information about the possibility 
that some symmetry may be broken spontaneously  \cite{KWetal}. No such terms 
were introduced here.

It is certainly possible to apply the reduction scheme described here to
other theories than Yukawa. Especially interesting case would be QCD, reduced 
to a nonrelativistic Schr\"odinger equation for bound states of constituent 
quarks, since such physical picture is the basis of phenomenology 
of hadrons in the particle data tables. It is not precisely clear how such 
picture could follow from the canonical QCD itself in any other approach.  
The fact found here that in the Yukawa theory one could enhance the 
coupling constant even up to 0.3 or 0.6 and the constituent picture remained 
convergent in the large-relative momentum region, is encouraging, because 
in QCD one expects the running coupling to be on the order of 1 for widths 
approaching 1 GeV, cf. \cite{GWafbs,Ggluon}.
  
\begin{figure}[htbp]
\begin{center}
\includegraphics[width=8cm]{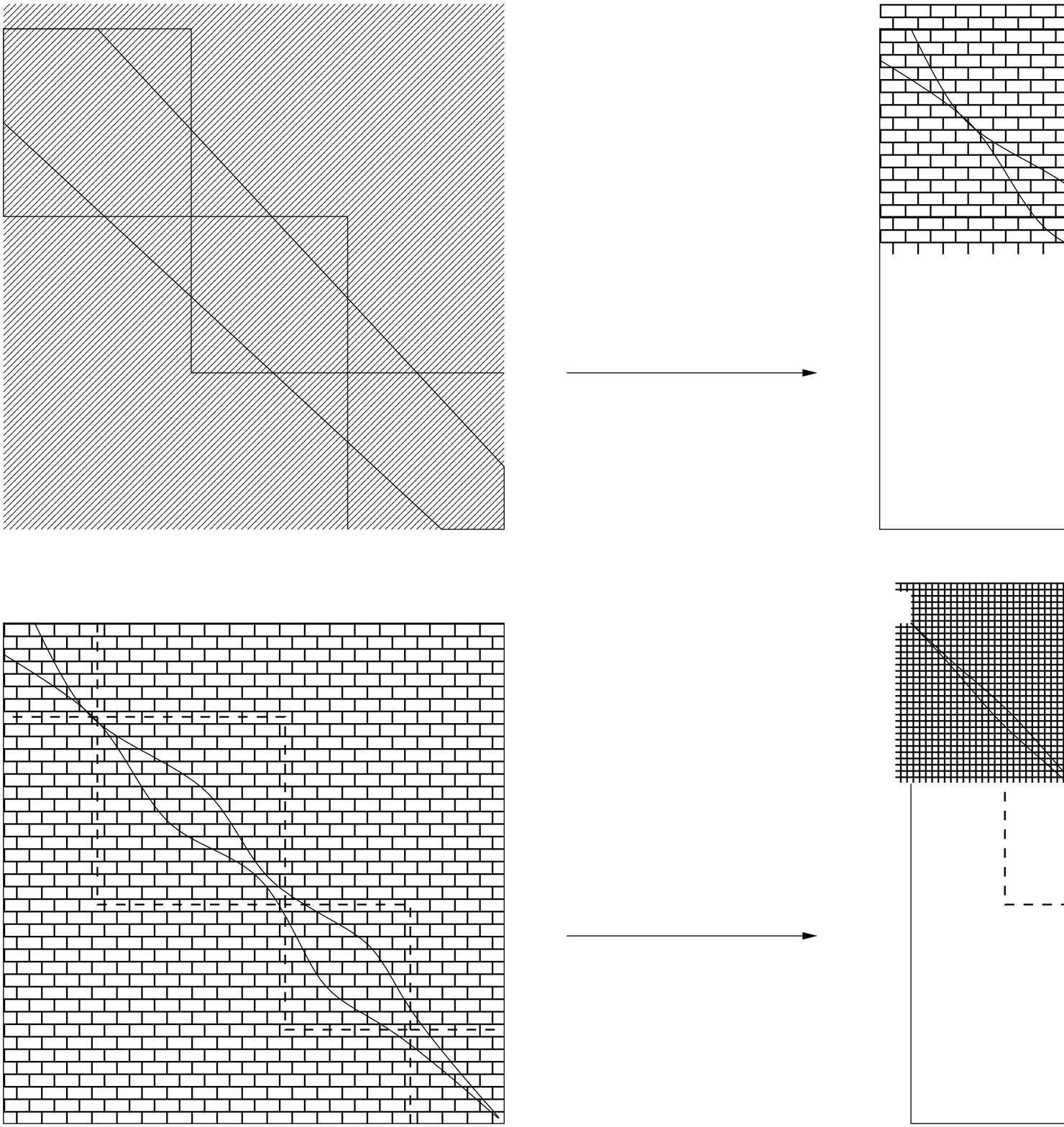}
\caption{Qualitative representation of the idea of solving the band diagonal
problem in effective Fock state basis in a sequence of diagonalizations of
``small'' matrices. The width of the shaded areas qualitatively represents
energy range of the couplings
between basis states obtained from a new diagonalization step.}
\label{fig:jelita}
\end{center}
\end{figure}

The discussion of bound states of effective fermions would not be complete 
without answering the question of how the binding of effective particles 
such as nucleons might emerge from a theory such as QCD. When the latter 
is already written in terms of effective quarks and gluons with small $\la$, 
the Hamiltonian matrix in the effective-particle basis can be cut into block 
pieces that lie along the diagonal, and touch each other with their corners 
(see Fig.~\ref{fig:jelita}), some matrix elements being left out.
Each of the small blocks can be treated as a separate theory and
solved. As a result, one obtains eigenstates or scattering states
which are closest to true solutions when their energies lie
somewhere in the middle of a block. The accuracy of the solutions
diminishes towards the corners where the small triangle parts should
provide increasingly important perturbations from couplings to the 
neighboring blocks. Fig.~\ref{fig:jelita} is drawn in the basis of 
effective quarks and gluons. Baryons and mesons in QCD would correspond 
to the approximate solutions in the individual blocks. It is natural 
to call the blocks channels and the calculation including the 
interactions between the blocks a coupled-channel theory. The 
transition from the original degrees of freedom (here, effective 
quarks and gluons) to the reduced degrees of freedom (baryons and 
mesons) is achieved by choosing the new basis of states: those given 
by the solution to the channels dynamics when they are treated as 
separate. Thus, the lowest mass channel with baryon number 2 contains 
only nucleons interacting through potentials, the next channel with
invariant masses between $2 m_N + m_\pi$ and $2m_N +2m_\pi$ may 
include two nucleons and one pion, etc. 
The bound or scattering states from higher blocks 
could be approximated by products of the eigenstates 
from lower-mass blocks. From the mathematical point 
of view, the transition step is a change of basis in describing 
eigenstates of relatively small matrices (in comparison to the 
initial Hamiltonian matrix) and taking into account the residual 
interactions between different channels later to refine the basis
states. The refinement step would be executed in the new basis using
the same type of division into blocks, but with the new blocks 
centered where the previous blocks were least accurate. This sequence 
of steps can be repeated many times. Thus, the whole transition process
to the effective theory written in terms of new degrees of freedom
would form a sequence of patterns indicated in Fig.~\ref{fig:jelita}.
The shaded shapes in Fig.~\ref{fig:jelita} indicate the
boundaries of non-zero matrix elements of the effective Hamiltonian
matrices  between the incoming and outgoing states, which 
qualitatively indicates their range in energy.

By the way, it is also natural 
to see in similar terms the transition from electrons and nuclei to 
atoms, or from atoms to molecules. The universal character of this 
scheme is the change of basis in the Hilbert space of states in 
quantum mechanics, and the Hamiltonian scheme for effective particles 
does open this option to detailed studies in all these theories. Once 
the width is sufficiently small, it slows down, or freezes degrees of 
freedom that require larger mass changes than $\la$ to be excited. 
One can also imagine that reducing the width farther in the new basis, 
a next level of effective degrees of freedom can be isolated using the 
same scheme. If the reduction were started with a theory underlying QCD,
i.e. at extremely large $\la$, the 
scope of the reduction scheme would be even greater.

As a final remark let us indicate that quantum field theories such as
Yukawa theory, or sigma-models, may be helpful in discovering the
acceptable structure of channel Hamiltonians even without making the
transition from QCD to nucleons and mesons explicitly.  This option 
follows from the fact that when the width in the effective theory 
is made small enough, the meson-fermion vertex contains a narrow
form factor and the emission of mesons is weak and slow, even for 
quite large coupling constants. The structure of the small-width 
effective theory may be then rather simple and universal in the
sense that many initial complex theories may reduce to the same small-width 
effective one. Thus, the apparently academic Yukawa theory, or other 
QFTs of that kind, may become very useful in indicating what kind 
of effective reduced Hamiltonians one can expect from the more elaborate 
analysis that derives nucleon-pion dynamics from QCD.  In fact, a small 
number of free parameters, such as the form factor width, the size of 
the coupling constant, or the size of effective masses in a basic set of
terms, may be sufficient 
to reproduce the bulk of the effective model predictions. It is also 
possible that no significant width dependence is observed in the model 
results if the parameters are correlated as if the effective model were 
derived from a Yukawa theory or a sigma-model using only a perturbative 
expansion for artificially small couplings that only later are extrapolated
to the required large values \cite{BGP,TM}.

\begin{acknowledgments}
This research has been supported in part by KBN grant number 2 PO3B 016 18.
\end{acknowledgments}
\appendix
\section{Reduction procedure}
\label{app:red}
The eigenvalue equation for a Hamiltonian $H$,
\begin{equation}
H\ket{\psi}=E\ket{\psi}\; 
\end{equation}
for some low eigenvalues, is replaced using the operation $R$ 
\cite{Bloch,Wold} by an eigenvalue equation for eigenstates  
$\ket{\varphi} = \sqrt{\hat P + R\hc R}\ket{\psi}$ of the
reduced Hamiltonian $H_R$, given by the following formula,
\begin{equation}
H_R = \frac{1}{\sqrt{1+R\hc R}}({\hat P}+R\hc)H({\hat
P}+R) \frac{1}{\sqrt{1+R\hc R}}\;.
\end{equation}
If one splits the initial Hamiltonian into the free and interaction
parts,
\begin{equation}
H=H_0+H_I\;,
\end{equation}
one can look for $R$ and $H_R$ in perturbation theory in $H_I$.  This
leads to the lowest (second order) expression for $H_R$:
\begin{equation}
\label{app:hr}
H_R={\hat P}H{\hat P}+\frac{1}{2} {\hat P}H_I{\hat
Q}\underline{H_I}{\hat P}- \frac{1}{2}{\hat P}\underline{H_I}{\hat
Q}H_I{\hat P}+\dots
\end{equation}
where the underlining denotes free energy denominators, displayed 
in Eq.(\ref{app:hrij}).

\section{Notation for spinors}
\label{app:spinors}
The four-component spinors of fermions and anti-fermions in
light-front Hamiltonians have the following form.
\begin{eqnarray}
u_{mp\lambda}&=&\frac{1}{\sqrt{mp^+}}\left[
p^+\Lambda_++\Lambda_-(m+\alpha^\perp p^\perp)\right] u_{\lambda}\;,
\nonumber \\
v_{mp\lambda}&=&\frac{1}{\sqrt{mp^+}}\left[
p^+\Lambda_++\Lambda_-(m+\alpha^\perp p^\perp)\right] v_\lambda\;,
\label{spinors}
\end{eqnarray}
where the projection operators are given by
\begin{equation}
\Lambda_\pm=\frac{1}{2}\gamma^0 \left(\gamma^0\pm\gamma^3\right)=
\frac{1}{2}\gamma^0 \gamma^\pm\;,
\end{equation}
the spinors of particles of zero velocity equal
\begin{equation}
\begin{array}{rclccrcl}
u_\uparrow&=&\sqrt{2m}\left(
\begin{array}{c}
\chi_1\\0
\end{array}
\right)\;&,&
u_\downarrow&=&\sqrt{2m}\left(\begin{array}{c}\chi_{-1}\\
0\end{array}\right)\;,\\
v_\uparrow&=&\sqrt{2m}\left(\begin{array}{c}0\\
\chi_{-1}\end{array}\right)\; &,&
v_\downarrow&=&\sqrt{2m}\left(\begin{array}{c}0\\-\chi_1\end{array}\right)\;,
\end{array}
\end{equation}
and the two-component spinors are
\begin{equation}
\chi_1=\left(\begin{array}{c}1\\0\end{array}\right)\quad \quad , \quad \quad
\chi_{-1}=\left(\begin{array}{c}0\\1\end{array}\right)\;.
\end{equation}
The matrices appearing on the right hand side of Eq. (\ref{spinors})
represent Lorentz transformations belonging to the small group
that preserves the four-vector $\eta$ defining the light front $x^+=0$
through the condition $\eta x = x^+$, where $x$ is an arbitrary four-vector.

Fermion field operator fulfilling free Euler-Lagrange equations $\psi_m(x)$ is
expanded for light-front ``time'' $x^+=0$ in terms of creation and annihilation
operators as: 
\begin{eqnarray}
&&\psi_m^{(i)}(x)|_{x^+=0}=\nonumber\\
&&=\sum_\sigma\int[p]\left[
b_{p\sigma}^{(i)}u_{mp\sigma}e^{-ip_\mu x^\mu}+
d_{p\sigma}^{(i)\dagger}v_{mp\sigma}e^{ip_\mu x^\mu}\right]\;.~~~~~~~
\end{eqnarray}

\section{Light-front momenta}
\label{app:xkapp}

For two particles of momenta $p_1$ and $p_2$, the
total momentum $P$ and the relative momenta $x$, $\kappa^\perp$ 
are defined as follows.
\begin{eqnarray}
p_1^+&=&xP^+\;,\\ 
p_2^+&=&(1-x) P^+\;,\\ 
p_1^\perp&=&xP^\perp+\kappa^\perp\;,\\
p_2^\perp&=&(1-x)P^\perp-\kappa^\perp\;.
\end{eqnarray}
Unless stated otherwise,
\begin{equation}
\delta^3(k-p)=\delta^2(k^\perp-p^\perp)\delta(k^+-p^+)\;.
\end{equation}
The shorthand notation in the integrals involves the measure
\begin{equation}
[p]=\frac{d^2p^\perp dp^+}{2 (2\pi)^3 p^+}\;.
\end{equation}
Creation and annihilation operators are normalized by the commutation relation
\begin{equation}
\left[a_p,a_k\hc\right]=2(2\pi)^3k^+\delta^3(k-p)\;
\end{equation}
for bosons, and in analogous way by anticommutation relations for each kind
of fermions.

In order to exhibit relationship between QFT 
and the well-known equal-time Schr\"odinger equation, one uses
relative momentum $\vec k$ \cite{jmn},
\begin{eqnarray}
k^\perp&=&\kappa^\perp\\ k_3&=&\sqrt{\frac{\kappa^{\perp
2}+m^2}{x(1-x)}}\left(x-\frac{1}{2}\right) \; ,
\end{eqnarray}
or, equivalently,
\begin{equation}
x=\frac{1}{2}\left(1+\frac{k_3}{\sqrt{\vec k^2+m^2}}\right)\;.
\end{equation}

\section{Schr{\"o}dinger's solution in
Coulomb potential}
\label{app:sch}
Equation
\begin{equation}
\frac{\vec k^2}{2\mu}\phi(\vec k)-\int \frac{d^3k'}{(2\pi)^3}
\frac{4\pi\alpha}{(\vec k- \vec k')^2} \phi(\vec k')=-B\phi(\vec k)
\end{equation}
(the Schr\"odinger equation for positronium without spin
and with reduced mass $\mu=m/2$) has the ground
state eigenvalue
\begin{equation}
\label{b0}
B_0=\frac{1}{2}\mu\alpha^2\;,
\end{equation}
and the normalized ground state wave function
\begin{equation}
\label{ff}
\phi_0(k)=N\frac{1}{(a^2+k^2)^2}\;,
\end{equation}
with $N=\sqrt{8\alpha^5\mu^5}/\pi$
and $a=\alpha \mu$.


\begin{thebibliography}{99}
\bibliographystyle{unsrt}

\bibitem{lattice}
K.~G.~Wilson, Phys.\ Rev.\ D {\bf 10}, 2445 (1974);
see also K.~G.~Wilson, in {\it C76-01-19.14} CLNS-327,
{\it presented at Coral Gables Conf., Miami, Fla, Jan 19-22, 1976};
and references therein.

\bibitem{SVZ}
M.~A.~Schifman, A.~I.~Vainstein, V.~I.~Zakharov, Nucl. Phys. B {\bf 147},
385 (1979); and references therein.

\bibitem{sim1} 
S.~D.~G\l azek and K.~G.~Wilson, 
%``Renormalization of Hamiltonians,'' 
Phys.\ Rev.\ D {\bf 48}, 5863 (1993).

\bibitem{KWetal} 
K.~G.~Wilson, et al., 
%``Non-perturbative QCD: A Weak coupling treatment on the light front,'' 
Phys.\ Rev.\ D {\bf 49}, 6720 (1994); 
and references therein.  
%[hep-th/9401153].

\bibitem{Gacta}  
S.~D.~G\l azek, 
%``Similarity renormalization group approach to boost invariant 
% Hamiltonian dynamics,''
Acta Phys.\ Polon.\ B {\bf 29}, 1979 (1998);  
%[hep-th/9712188].  
{\it ibid.}
%``Running couplings in Hamiltonians,'' %Acta Phys.\ Polon.\  B
{\bf 31}, 909 (2000).  
%[arXiv:hep-th/0001042].  %%CITATION = HEP-TH 0001042;%%

\bibitem{GWafbs} 
S.~D.~G\l azek and K.~G.~Wilson, 
%``Asymptotic freedom and bound states in Hamiltonian dynamics,'' 
Phys.\ Rev.\ D {\bf 57}, 3558 (1998).  
%[hep-th/9707028].

\bibitem{Wegner}  
	F. Wegner, 
	Ann. Phys. (Leipzig) 3, 77 (1994).

\bibitem{Ggluon} 
	S.~D.~G\l azek,  
	% ``Dynamics of effective gluons''
	Phys.\ Rev.\ D {\bf 63}, 116006 (2001).

\bibitem{poincare} 
	S.~D.~G\l azek, T. Mas\l owski,
	%``Renormalized Poincar\'e algebra for effective particles in quantum 
	% field theory'' 
	Phys.\ Rev.\ D {\bf 65}, 065011 (2002).

\bibitem{WPR}
        F. J. Wegner, 
        Phys.\ Rep.\ {\bf 348}, 77 (2001).

\bibitem{PPR}
        R. J. Perry,
        Phys.\ Rep.\ {\bf 348}, 33 (2001); and references therein.       

\bibitem{tamm}
	I.~Tamm,
	%``Relativistic Interaction Of Elementary Particles,''
	J.\ Phys.\ (USSR){\bf 9}, 449 (1945).

\bibitem{dancoff}%{Dancoff}
	S.~M.~Dancoff,
	%``Nonadiabatic Meson Theory Of Nuclear Forces,''
	Phys.\ Rev.\ {\bf 78}, 382 (1950).

\bibitem{pinsky}
	S.~G\l azek, A.~Harindranath, S.~Pinsky, J.~Shigemitsu and K.~Wilson,
	%``On the relativistic bound state problem in the light front 
	% Yukawa model,''
	Phys.\ Rev.\ D {\bf 47}, 1599 (1993).

\bibitem{Perry:mz}
        R.~J.~Perry, A.~Harindranath and K.~G.~Wilson,
        %``Light Front Tamm-Dancoff Field Theory,''
        Phys.\ Rev.\ Lett.\  {\bf 65}, 2959 (1990).
        %%CITATION = PRLTA,65,2959;%%

\bibitem{zle}%10.7
	%Pauli Villars:
	S.~J.~Brodsky, J.~R.~Hiller and G.~McCartor,
	%``Application of Pauli-Villars regularization and 
	% discretized light-cone  quantization to a (3+1)-dimensional model,''
	Phys.\ Rev.\ D {\bf 60}, 054506 (1999)
	%[hep-ph/9903388].
;
	S.~J.~Brodsky, J.~R.~Hiller and G.~McCartor,
	%``Pauli-Villars as a nonperturbative ultraviolet regulator in 
	%discretized  light-cone quantization,''
	Phys.\ Rev.\ D {\bf 58}, 025005 (1998)
	%[hep-th/9802120]
;       
        H. C. Pauli, S. J. Brodsky, Phys.\ Rev.\ D {\bf 32}, 2001 (1985). 

\bibitem{casher}
	A.~Casher, Phys.\ Rev.\ D{\bf 14}, 452 (1976);
	C.B.~Thorn, Phys.\ Rev.\ D{\bf 20}, 1934 (1979);
	cf. H.~C.~Pauli, S.~J.~Brodsky, 
	Phys.\ Rev.\ D{\bf 32}, 2001 (1985). 


\bibitem{BP}
        M.~Brisudova and R.~J.~Perry, 
	% ``Initial studies of bound state in light-front QCD''
	Phys.\ Rev.\ D {\bf 54}, 1831 (1996); 
	%M.~Brisudova and R. J.~Perry
	% ``Note on restoring manifest rotational symmetry in hyperfine and
	% fine structure in light-front QCD''
	{\it ibid.} D {\bf 54}, 6453 (1996);
        M. Brisudova, R. J. Perry, K. G. Wilson, 
	Phys. Rev. Lett. {\bf 78}, 1227 (1997).

\bibitem{Jones}
	B.~D.~Jones, R.~J.~Perry and S.~D.~G\l azek,
	%``Nonperturbative QED: An analytical treatment on the light front,''
	Phys.\ Rev.\ D {\bf 55}, 6561 (1997).
	%[hep-th/9605231].

\bibitem{briprep}
M.~Brisudova,
%``Small x divergences in a heavy quark-antiquark state,''
Nucl.\ Phys.\ Proc.\ Suppl.\  {\bf 108}, 173 (2002); 
%[arXiv:hep-ph/0111388].
%%CITATION = HEP-PH 0111388;%%
M.~Brisudova,
%``Small x divergences in the similarity RG approach to LF QCD,''
Mod.\ Phys.\ Lett.\ A {\bf 17}, 59 (2002).
%[arXiv:hep-ph/0111189].
%%CITATION = HEP-PH 0111189;%%
\bibitem{Bloch}
        C. Bloch, Nucl. Phys. {\bf 6}, 329 (1958).

\bibitem{Wold}
	K. G.~Wilson
	Phys.\ Rev.\ D{\bf 2}, 1438 (1970).

\bibitem{Yan}
        See e.g. 
        K. Bardakci and M. B. Halpern, Phys. Rev. {\bf 176}, 1686 (1968);
        L. Susskind, {\it ibid.} {\bf 165}, 1535 (1968);
        S.-J. Chang, R. G. Root, and T.-M. Yan, {\it ibid.} 
	D{\bf 7}, 1133 (1973);
        S.-J. Chang and T.-M. Yan, {\it ibid.}, 1147; T.-M. Yan, {\it ibid.}, 
	1760, 1780, and references therein. 

\bibitem{Glazek1980}
	S.~G\l azek, ``Weinberg equation with spin'', M. Sc. diploma work,
        Warsaw University, Institute of Theoretical Physics (1980); see also
        Acta Phys.\ Polon.\ B {\bf 15}, 889 (1984).

\bibitem{karmanov}
	M.~Mangin-Brinet, J.~Carbonell and V.A.~Karmanov,
%	hep-th/0102068
	Phys.\ Rev.\ D{\bf 64}:125005 (2001).
%	[hep-th/0107235]

\bibitem{overlap}
	S.~D.~G\l azek and K.~G.~Wilson,
	%``Renormalization of overlapping transverse divergences in a 
	%model light front Hamiltonian,''
	Phys.\ Rev.\ D {\bf 47}, 4657 (1993).

\bibitem{BGP}
	A.~B.~Bylev, S.~D.~G\l azek and J.~Przeszowski,
	%``Relativistic Hamiltonian dynamics of pions in nucleons,''
	Phys.\ Rev.\ C {\bf 53} (1996) 3097.

\bibitem{TM}
	T.~Mas\l owski and M.~Wi\c eckowski,
	%``4th order similarity renormalization of a model hamiltonian,''
	Phys.\ Rev.\ D {\bf 57}, 4976 (1998).
	%[hep-th/9707057].

\bibitem{jmn}
	See e.g. P.~Danielewicz and J.~M.~Namys\l owski, 
	Phys.\ Lett.\ {\bf 81B} 110 (1979).

\end{thebibliography}
\end{document}